 \documentclass[final,3p,times,twocolumn]{elsarticle}
\usepackage{amssymb}
 \usepackage{lineno}

\journal{Astroparticle Physics}
\begin{document}

\begin{frontmatter}

%% Title, authors and addresses
%% use the tnoteref command within \title for footnotes;
%% use the tnotetext command for the associated footnote;
%% use the fnref command within \author or \address for footnotes;
%% use the fntext command for the associated footnote;
%% use the corref command within \author for corresponding author footnotes;
%% use the cortext command for the associated footnote;
%% use the ead command for the email address,
%% and the form \ead[url] for the home page:
%%
%% \tnotetext[label1]{}
%% \author{Name\corref{cor1}\fnref{label2}}
%% \ead{email address}
%% \ead[url]{home page}
%% \fntext[label2]{}
%% \cortext[cor1]{}
%% \address{Address\fnref{label3}}
%% \fntext[label3]{}    	
\title{
{\small\mbox{}\hfill DESY 11-059}\\[1.5ex]
A method for   untriggered time-dependent searches  for  multiple flares from  neutrino point sources}
%% use optional labels to link authors explicitly to addresses:
%% \author[label1,label2]{<author name>}
%% \address[label1]{<address>}
%% \address[label2]{<address>}

\author[a,b]{D. G\'ora}
\author[a]{E.~Bernardini}
\author[a]{A.H. Cruz Silva}

\address[a]{DESY, D-15735 Zeuthen, Germany}
\address[b]{Institute of Nuclear Physics PAN, ul. Radzikowskiego 152, 31-342 Cracow, Poland}

\begin{abstract}
%% Text of abstract
A method for a time-dependent search for flaring  astrophysical 
sources which can be potentially detected by large neutrino experiments is presented.
The method uses a time-clustering algorithm combined with an unbinned likelihood procedure. 
By including in the  likelihood  function a signal term  which describes  the
contribution of  many  small clusters of signal-like events, this method provides
an effective way for  looking for   weak  neutrino flares  over
different time-scales. The method is sensitive to an overall excess of events
distributed over several flares which are not individually detectable. For standard cases (one flare)
 the  discovery potential  of  the method is  worse than a standard time-dependent point source analysis with unknown duration of the flare by a factor depending on the signal-to-background level.
However, for flares sufficiently shorter than the total observation period,  the  method is more sensitive  than a time-integrated analysis.

% When the number of individual flares in  analyzed data period  is increased the number of  events needed for discovery decreases
% comparing to one flare with the similar duration, especially in the case of a few weak flares distributed %over  a  longer period.
\end{abstract}

\begin{keyword}
Neutrino, Time-dependent searches, Neutrino experiments
%% keywords here, in the form: keyword \sep keyword

%% MSC codes here, in the form: \MSC code \sep code
%% or \MSC[2008] code \sep code (2000 is the default)

\end{keyword}

\end{frontmatter}

%%
%% Start line numbering here if you want
%%
% \linenumbers
\section{Introduction} 
\label{sec1}
% If the  astrophysical sources of extragalactic neutrinos are flaring,  their detection probability
%   is enhanced  by online searches for correlations with established signals (e.g. flares in high-energy gamma-rays)
% by neutrino observations.  For example, blazars, among all Active Galactic Nuclei (AGNs), 
% show the most extreme photon flux variability at all wavelengths. The variability time-scale ranges from fast flares of few minutes, to hours 
% and high states of several days. In hadronic models of AGNs ~[\cite{Aharonian}], these flares arise from photo-hadronic interactions, 
% which also lead to the coincident production of neutrinos. Transient sources, like Gamma-Ray Bursts (GRBs), 
% are also candidates of photon-correlated astrophysical neutrinos. Their shorter duration, in the order of seconds, makes their search almost background free. 

The major aim of neutrino astrophysics is to contribute to the understanding of the origin of high energy cosmic rays. A point-like neutrino signal of cosmic origin would be an unambiguous signature of hadronic processes, unlike $\gamma$-rays which can also be created in leptonic processes. Neutrino telescopes are ideal instruments to monitor the sky and look for the origin of cosmic rays because they can be continuously operated. The detection of cosmic neutrinos is however very challenging because of their small interaction cross-section and because of a large  background of atmospheric neutrinos. Parallel measurements using neutrino and electromagnetic observations (the so-called "multi-messenger" approach) can increase the chance to discover the first neutrino signals by reducing the trial factor penalty arising from observation of multiple sky regions and over different time periods. In a longer term perspective, the multi-messenger approach also aims at providing a scheme for a phenomenological interpretation of the first possible detections.

The search of occasional flares with a high-energy neutrino telescope is motivated by the high variability which characterizes the electromagnetic emission of many neutrino candidate sources. Resent results obtained by the IceCube Collaboration~\cite{Bazo} indicate that high-energy neutrino telescopes have reached a sensitivity to neutrino fluxes which is comparable to the observed high energy gamma-ray fluxes of Blazars in the brightest states (e.g. the flares of Markarian 501 in 1997~\cite{aha1} and Markarian 421 in 2000/2001~\cite{aha2}). With the assumption that the possibly associated neutrino emission would be characterized by a flux enhancement comparable to what is observed in gamma-rays in such states, neutrino flares could be extracted from the sample of neutrino-like events with a reasonable significance. 

These astrophysical neutrinos can be searched for in several ways. One of the methods 
for  a neutrino point source search  is to  look for events coming from a restricted angular region, 
which could be identified with a known astrophysical object.  Finding neutrino point sources  in the sky means to locate 
an excess of events from a particular direction over the background  of  atmospheric neutrinos and muons. 
 These events might present additional features that distinguish them from background, for example a different energy spectrum or time structure.
For  sources which manifest large time variations in the emitted electromagnetic radiation,  the signal-to-noise ratio can be increased by testing smaller time windows  around the flare (a  time-dependent  search). In principle there are two approaches to neutrino time-dependent searches: 
\begin{itemize}
\item \textbf{Triggered Flare Search}: Looking directly for photon-neutrino correlations 
using specific source lightcurves from Multi-WaveLength (MWL) observations~\cite{Baker}.
\item \textbf{Untriggered Flare Search}: A generalized search (MWL data)  for neutrino flares, motivated but not associated with MWL observations, which are scarce and not available for all sources during complete periods. In addition, there could be a time lag between observed photon flares and the associated neutrino
 flares. In the extreme cases high energy photons could be entirely  absorbed 
 during periods of  the  highest photon production in the source~\cite{Dermer}.  
This approach however entails a higher trial factor penalty than  triggered flare searches. As a merit, neutrino flares which are not accompanied by 
 observed electromagnetic counterparts are not automatically excluded. This approach is also less dependent on models for the correlation between the neutrino and the electromagnetic emission and not dependent on the availability of multi-wavelength information. 
\end{itemize}

Here we develop a method that is well suited for an intermediate approach in which ``periods of interest''
can be a priori selected on the basis of  multiwavelength data. The neutrino sample can 
then be scanned looking for  significant structures, in a way which is less dependent on models   predicting
different correlations with  a given wavelength.

An untriggered unbinned flare search was    first developed and applied to 
IceCube data, using a compact list of pre-defined source directions~\cite{Bazo}. A time-clustering algorithm  \cite{Bazo,Satalecka}, and the   unbinned maximum likelihood method~\cite{Braun,Braun2} are the basis of this analysis.  Such a  method  finds the most significant flare 
in a long period. The number of trials coming from all combinations of event times is increased, reducing the significance. However, for flares sufficiently shorter than the total observation period, 
the time clustering algorithm is more sensitive  than a time-integrated analysis. 

In this paper,  we  propose  an extension of the  method   described   in~\cite{Bazo}.  By including in the  likelihood  a signal term  which describes  the contribution of  many  small clusters  of signal-like events, our algorithm can extract  not only  the most significant  flare,
 but also less significant clusters of events   distributed over  weak flares.   These  weaker flares could
be separated  by any distance in time  and   will be very difficult to detect  or even   can not be   detected with  other  methods~\cite{Braun,Braun2}.

The paper is  structured as   follows. The algorithm is described in Section 2. In Section~3  we  apply
 the method   to a simulated  neutrino search   for one  flare   and a few weak flares  separated in time. 
A  short  discussion  about the algorithm  performance  is presented in Section 4.  Conclusions are given in Section 5.

\section{The algorithm }
In this section, we first    describe  the time-clustering algorithm, then   
we  shortly   recall the unbinned maximum likelihood method~\cite{Braun}
and  finally   we  propose  our new  algorithm. 

\begin{figure}[t]
 \centering
 \includegraphics[width=0.45\textwidth]{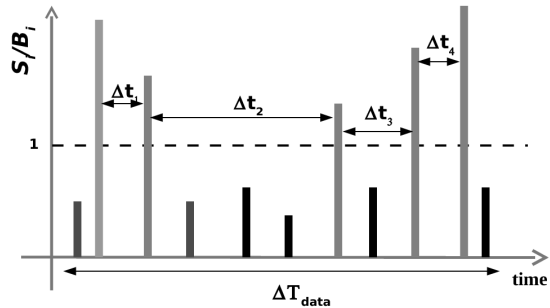}
 \caption{\small The basic  idea of the time clustering procedure. The signal-like events are extracted out of a data sample and then sorted in time. Each combination of such event pairs defines a possible flare time window, which is  then tested with a maximum likelihood method.}\label{figure1}
 \end{figure}

\subsection{The  time-clustering algorithm}
The time clustering algorithm~\cite{Bazo}  selects the most significant cluster of events in time  
and returns the mean time and width of the corresponding flare. The basic idea
 is  shown in Figure~\ref{figure1}.  In a first step, the method selects the most promising flare candidates over different  time windows ($\Delta t_j$), which are given  by the combination of the times of
 signal-like events from the analyzed data set. A signal-like event is defined  as having
$S_{i}/B_{i}>1$, where $S_{i}$ and $B_{i}$ is  the background and  the signal Probability 
Density Function (PDF)   as defined for  the time-integrated method~\cite{Braun}\footnote{ 
To calculate the  ratio, $S_{i}/B_{i}>1$,   only the spatial and energy terms in the PDF's are included.}.
  Each combinations of these event times defines the start 
and end time ($t^{\mathrm{min}}_j$ and $t^{\mathrm{max}}_j$) of a candidate  flare  time window ($\Delta t_{j}$).
For each $\Delta t_j$, a significance parameter (the test statistic) $\mathrm{TS}_j$ is then calculated as defined in~\cite{ Braun}. Larger values of $\mathrm{TS}_j$   correspond to data  less compatible 
with the null  hypothesis (i.e. zero expected signal events in the data sample tested).
Finally,  the algorithm returns the best $\mathrm{TS}_j$, $\mathrm{TS}_{\mathrm{max}}$, corresponding to the most significant time window over the entire data  period analyzed.

\subsection{The unbinned maximum likelihood method}
The unbinned maximum likelihood method~\cite{Braun} defines the test statistic parameter by: 
\begin{equation}
  \mathrm{TS}_{j}=-2 \log\left[ \frac{\mathcal{L}(\vec{x}_{s},n_s=0)}{\mathcal{L}(\vec{x}_{s},\hat{n}_s,\hat{\gamma}_s)}  \right],
\label{lambda2}
\end{equation}
where $\vec{x}_{s}$ is the source location, $\hat{n}_s$ and
$\hat{\gamma}_s$ are the best estimates of the number of signal events
and source spectral index, respectively, which are found by maximizing
the likelihood  $\mathcal{L}$:

\begin{equation}
  {\mathcal L}(n_s,\gamma,\Delta t_{j})=\prod_{i=1}^{N} \left ( \frac{n_{s}}{N}S_{i}+ \left( 1-\frac{n_{s}}{N} \right) B_i \right ),
 \label{llh_unbinned}
\end{equation}
where $N$ is the number of all events in  the  data sample.

The  background  PDF  is given by:
\begin{equation}
B_{i}=P^{\mathrm{space}}_{i}(\theta_{i},\phi_{i})P^{\mathrm{energy}}_{i}(E_{i},\theta_{i})P^{\mathrm{time}}_{i}(t_{i},\theta_{i}),
\label{llh_bg_time}
\end{equation}
where $P^{\mathrm{space}}_{i}$ describes the space distribution of events  at a given region of the sky  with the solid  angle $d\Omega$, $P^{\mathrm{energy}}_i$ is the energy  distribution 
and $P^{\mathrm{time}}_{i}$ describes the background time  distribution. These PDFs can be calculated purely from data. In general, due to  analysis cuts, Earth absorption effects and the detector geometry, the spatial probability $P^{space}_{i}$ and the energy probability  $P^{\mathrm{energy}}_i$  depend on zenith, $\theta_i$, and azimuth, $\phi_i$. 

The time probability $P^{\mathrm{time}}$ is defined by:
\begin{equation}
P^{\mathrm{time}}_{i}(t_{i},\theta_{i})=\frac{1}{\Delta T_{\mathrm{data}}},
\label{bg_time}
\end{equation}
where $\Delta T_{\mathrm{data}}$ is a normalization constant for the whole data taking period.
 
The properties of signal events are taken from a dedicated  Monte Carlo (MC) signal simulations.
 The signal PDF, $S_i$   is given by:
\begin{equation}
S_{i}=P^{\mathrm{space}}_i(\mid \vec{x}_i-\vec{x}_{s} \mid, \sigma_{i})P^{\mathrm{energy}}_i(E_{i},\theta_{i},\gamma_{s})P^{\mathrm{time}}_{i},
\label{llh_signal_time}
\end{equation}
where the spatial probability $P^{\mathrm{space}}_i$ is a Gaussian function of $\mid\vec{x}_i-\vec{x}_{s}\mid$, the space angular difference between the source location $\vec{x}_{s}$ and each event's reconstructed direction $\vec{x}_{i}$, and $\sigma_{i}$ the angular error estimate of the reconstructed track. The energy probability $P^{\mathrm{energy}}_i$, constructed from signal simulation, is a function of the event energy estimator $E_i$, the zenith coordinate $\theta_i$, and the assumed energy spectral index of the source $\gamma_s$ (such that $\frac{dN}{dE}\sim E^{-\gamma_s}$).  $P^{\mathrm{time}}_{i}$  is the  time probability which generally is  
a constant value (i.e. taken to be uniform in time) if no flare structure is assumed. For each 
time window tested (${\Delta t_j}=t^{\mathrm{max}}_j-t^{\mathrm{min}}_j$), the time probability is  given  by:
\begin{equation}
 P^{\mathrm{time}}=\frac{H(t^{\mathrm{max}}_j-t_i) H(t_i-t^{\mathrm{min}}_j)}{\Delta t_j},
\end{equation}
where $t_i$ is the arrival time of the $i^{\mathrm{th}}$ event and  $H$ is the Heavyside step function. 
Note, that by using  this definition for  the time probability   in the  likelihood $\mathcal{L}$, we count
 only  events     which fall inside the current  time window ${\Delta t_j}$ (i.e. the signal PDF  $S_i$ is zero outside  the  selected time window).

As  shown in~\cite{Braun2}  the unbinned likelihood  method 
will preferentially find  shorter flares, making it less powerful for flares of durations longer than roughly one day. The solution to this problem is to use
 a  marginalization term $\left(\frac{\Delta t_j}{ \Delta T_{\mathrm{data}}}\right)$ in the likelihood $\mathcal{L}$~\cite{Braun2}. This 
 gives a more uniform exposure to find flares of different   widths and  leads to  
a redefinition of the test statistic: 
\begin{equation}
  \mathrm{TS}_{j}|_{\Delta t_j}=-2 \log\left[\frac{\Delta T_{\mathrm{data}}}{\Delta t_j} \times \frac{\mathcal{L}(\vec{x}_{s},n_s=0)}{\mathcal{L}(\vec{x}_{s},\hat{n}_s,\hat{\gamma}_s)}  \right].
\label{lambda3}
\end{equation}

\subsection{A method to search for multiple flares}
 
We propose here some extensions to the above mentioned procedures
to  identify a series of weak flares   by incorporating this  
information into  the likelihood function.  This is done by restricting  the search  to  
doublets  of  signal-like events, and using the value of the  test-statistic 
of  these individual flare candidates as their weights in a stacking-like calculation 
of the global maximum likelihood.

\begin{figure}[t]
 \centering
\includegraphics[width=0.45\textwidth]{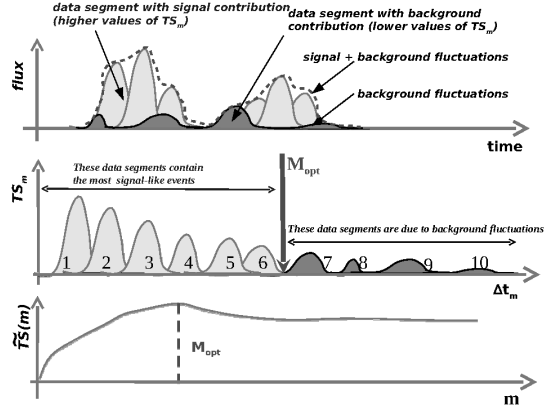}
 \caption{\small Sketch of the flare  stacking procedure. Top: A data set is divided into segments being defined as in Figure~\ref{figure1}. Some of these segment will include signal contribution, others background fluctuations only; Middle: Sorted values of the test statistic for each data segment from a maximization of the likelihood in Eq.~(2) as a function of the data segment index $m$. 
 Bottom: Evolution of the test statistic from a maximization of the likelihood in Eq.~(8) as a function of the number of data segments being stacked following Eq.~(9). The maximum value of this parameters defines the optimal number of data segments to be stacked $M_{\mathrm{opt}}$.}\label{figure2}
 \end{figure}

First, we   extract all doublets  that can be formed out of all 
signal-like events ($S_i/B_i>1$)  over the entire data taking period $\Delta T_{\mathrm{data}}$.
This step serves to isolate all possible (and smallest) time  windows 
that compose the signal contribution in the tested data sample (the total number being $M$).
 We call these time windows ``data segments''. Note,  that by construction, there 
could be  a certain degree of overlap between different data segments if larger
event  multiplicity will be studied, see Figure~\ref{figure2} Top.
The choice of doublets is  physics-motivated, because for a  neutrino  detector
like IceCube,  the signal expectation is not much more than a few  signal neutrinos per year  from  the  strongest
astrophysical sources~\cite{time}.  Moreover, we focus on weak multiple flares.

Then for each time window $\Delta t_j$   a minimization of $\log(\mathcal{L})$  as defined in Eq.~(2)
is performed, with $n_{s}$ and $\gamma_{s}$ as free parameters, and the individual test statistic  $\mathrm{TS}_{j}|_{\Delta t_j}$ is calculated. This step serves to estimate the possible signal contribution in each data segment. All  time windows are then  sorted  according to  $\mathrm{TS}_{j}|_{\Delta t_j}$,  as it is  shown in  Figure~\ref{figure2} (Middle). 
 Some of these data segments  will contain  
real  signal events  and some of them  are  likely due to  background fluctuations. 
Our aim is to extract   the optimal (best suited) number of  data segments ($M_{\mathrm{opt}}$) 
which compose the total signal contribution injected  in the overall period  $\Delta T_{\mathrm{data}}$.

For this purpose,  we propose a  modification of the single-source  likelihood function (Eq.(2))
 by  including a signal term ($S_{i}^{tot}(m)$) describing  the contribution of each data segments:
\begin{equation}
 \tilde{{\mathcal L}}(n_s,\gamma,m)=\prod_{i=1}^{N} \left ( \frac{n_{s}}{N}S_{i}^{\mathrm{tot}}(m)+ \left( 1-\frac{n_{s}}{N} \right) B_i \right )
 \label{llh_unbinned2}
\end{equation}
where $m$ is the number of   data segments in to which the overall signal contribution can be decomposed  (a free parameter $m \leq M$)
and $N$ is the total number of events in the time period $\Delta T_{\mathrm{data}}$.

In  other words,  in order to include  the
contribution  to the signal  from  multiple flares, the one-source signal 
term $S_{i}$  is being replaced by   the sum of signal  sub-terms over $m$ 
data-segments:
\begin{equation}
S_{i} \rightarrow S_{i}^{\mathrm{tot}}(m)=\frac{\sum_{j=1}^{m} W^j \times S_i^{j}(\mid \vec{x}_i-\vec{x}_{s} \mid, E_i,\gamma,\Delta t_j) }{\sum_{j=1}^{m} W^j}
 \label{llh_unbinned3}
\end{equation}
where  $W^{j}$ is a weight which  describes the strength (significance) of the doublet
contained in each   data segment.
This weight   should  be proportional  to the expected number of neutrino  events  
seen in  the detector.  As we will  show later 
the  test statistic  is quite well correlated
with the  true number of  injected signal events. Thus  we  take   $W^{j}=\mathrm{TS}_{j}|_{\Delta t_j}$. 
 
The  likelihood  given in Eq.~(\ref{llh_unbinned2}) in combination with Eq.~(9)
 is  usually  used in  stacking searches
for different types of  astrophysical point sources~\cite{IC_GRB}, 
but  here the individual  sources   are data segments of the signal-like 
events found in the data from the same object. The stacking  method 
uses the fact  that signal and background increase differently when observations
for multiple sources (or flares) are being added.
% While the signal grows linearly with the number of $m$ sources the background grows proportionally
% to $\sqrt{m}$. 
Stacking multiple flares can therefore suppress the overall
 background and hence can increase  the signal sensitivity.

In order to estimate the optimal number of data segments  $M_{\mathrm{opt}}$
for a given  number of $m$ segments (starting from $m=1$) we minimize the  
 $\log(\tilde{{\mathcal L}}(n_s,\gamma,m))$ 
with  $n_s$ and  $\gamma_s$ as  free parameters. The  minimization   returns the best estimates for the 
number of  signal events  $\hat{n}_{s}$  and for the spectral index of the source $\hat\gamma_s$, 
and  the  ``global'' test statistic  is calculated  from:

\begin{equation}
 \widetilde{\mathrm{TS}}(m)=-2 \log\left[ \frac{\tilde{\mathcal{L}}(\vec{x}_{s},n_s=0)}{\tilde{\mathcal{L}}(\vec{x}_{s},\hat{n}_s,\hat{\gamma}_s,m)}  \right].
\label{lambda}
\end{equation}

Then,  the optimal  number of  data segments to be stacked  ($M_{\mathrm{opt}}$)
 is chosen according to the maximum  of $\widetilde{\mathrm{TS}}(m)$ (see Figure~\ref{figure2} Bottom). 
As a result,  the following important parameters  are  extracted:
\begin{itemize}
 \item $M_{\mathrm{opt}}$: the optimal number of segments which compose the signal contribution over the time 
$\Delta T_{\mathrm{data}}$. The optimal set   could be either subsequent in time, forming one   significant cluster of events (one flare) for a given source location, 
or  theses segments can be distributed  among  a  few (sometimes less significant) flares
 separated in time. This  is shown  in  Figure~\ref{figure2}~(Top).
 \item $\hat{n}_{s}$: the total number of expected signal events summed over the  $M_{\mathrm{opt}}$ individual segments.
 \item $\hat{\gamma}_{s}$: the spectral index of the source (the flare), assumed to be 
the same   for each $M_{\mathrm{opt}}$ segments.
 \item $\widetilde{\mathrm{TS}}(M_{\mathrm{opt}})$: the maximum value of the  test statistic calculated with  the modified 
likelihood function of Eq.~(\ref{lambda}).
\item $\Delta T(M_{\mathrm{opt}})$: the flare duration  calculated  for  $M_{\mathrm{opt}}$ segments. 
This is here defined 
as the time between the start  time of the first  data segment  and the end time of the last data segment.
\end{itemize}

The overall significance of the optimal  configuration $M_{\mathrm{opt}}$ can be determined
using MC simulations by applying the same procedure to a large
number of simulated  data samples. This will automatically
 account for effects of trial factors.  The trial factors arise from testing different time windows for the same source direction. In order to represent a background-only observation, the properties of the data events (e.g. zenith, azimuth, time, reconstruction error and energy estimator) are sampled from their distributions. Data of  neutrino experiments with a signal flux can be simulated
 by injecting signal events on top of the background data.

 \section{Results}
 
The  method described in Section~2   has been applied to  10,000  MC background  samples
(scrambled sky  maps). Background events were randomly chosen reproducing realistic  PDF's 
(energy, angular resolution)  extracted from~\cite{time}. More precisely,
 the number of background events in a given declination band ($n_{\mathrm{bg}}^{\mathrm{band}}$) is estimated  from $n_{\mathrm{bg}}^{\mathrm{band}}=n_{\mathrm{bg}}\times \frac{ A_{\mathrm{
band}}} {A_{\mathrm{bin}}} \times \frac{\Delta T_{\mathrm{data}}}{375.5} $,  where  $n_{\mathrm{bg}}$ is  the  number of background events  in a  bin  and  $ A_{\mathrm{band}}$ and $A_{\mathrm{bin}}$ are the solid angles of the band and bin, respectively. Assuming $n_{\mathrm{bg}}=2.5$  background events  for    declination $\delta=15^{\circ}$ as in~\cite{time} and $\Delta T_{\mathrm{data}}=40$ days we  are left with 355 events  distributed
in a  declination band of size  $\pm6^{\circ}$. The angular reconstruction  error $\sigma_{i}$ 
of each event  was  generated   based on the cumulative spread function   from~\cite{time} with a median  resolution of about $0.8^{\circ}$ and  assuming that $\sigma_i$  does not depend  on the energy 
of the reconstructed event. This assumption in general may not be  true, but it won't  affect the 
 results of this test. The same number of scrambled maps were  generated for  signal events injected on  top of  the background (MC background plus signal simulations). 
A neutrino point  source  at declination  $15^{\circ}$  was considered.
 Signal events were injected   around   this  declination  $\delta_{s}$ along a band  
 $\delta_{s}-6^{\circ} <\delta_{s}< \delta_{s}+6^{\circ}$~\footnote{Events outside this band hardly contribute to the signal PDF since they are too far  away from the source compared to the point spread function of~\cite{time}.}
and with azimuth   randomly chosen  from $0^{\circ}$ to $360^{\circ}$. 
Individual event directions were smeared according to  the Point Spread Function (PSF) from~\cite{time}  with a  median angular resolution  of about $0.8^{\circ}$. 
 We  simulate    signal strengths following   Poisson  statistics with   mean values of 4, 8, 9 and  12 signal  events. The energy PDF follows what 
was found  for  an  $E^{-2}$  energy spectrum in~\cite{time}.
For each MC signal  sample (scrambled sky map),  signal events were randomly injected inside 
a time window defined  by  various flare durations, in the range from 0.01 day to 30 days. The most of the results presented in this paper was obtained   considering  a total period of
$\Delta T_{\mathrm{data}}=40$ days. This is  motivated by the fact  that an obvious application of this method is to test periods of interests, 
preselected with the help of  multi-wavelengths data.  

\begin{figure}[t]
 \centering
 \includegraphics[width=0.45\textwidth]{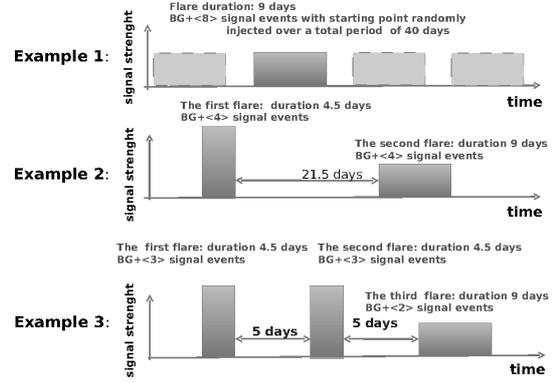}
 \caption{\small Sketch  of the   three examples discussed  in this work i.e. one flare (Example~1) 
two individual flares  (Example~2)  and three weak flares separated in time (Example~3).}\label{figure3}
 \end{figure}
We tested the method  for  the   three  following   cases which are also  illustrated in Figure~\ref{figure3}.
\begin{figure*}[ht!]
 \centering
 \includegraphics[width=7 cm,height=5cm]{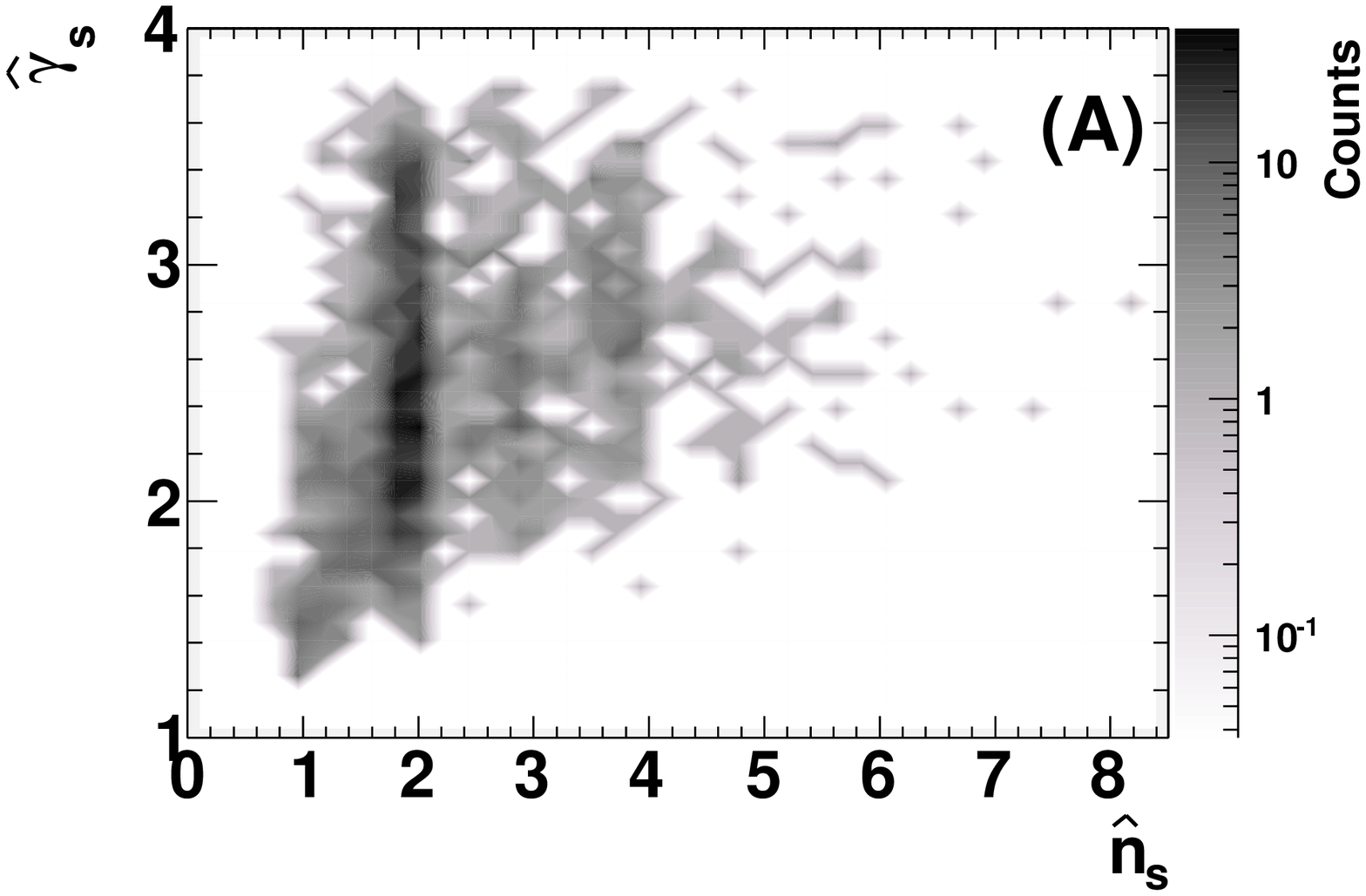}
 \includegraphics[width=7 cm,height=5cm]{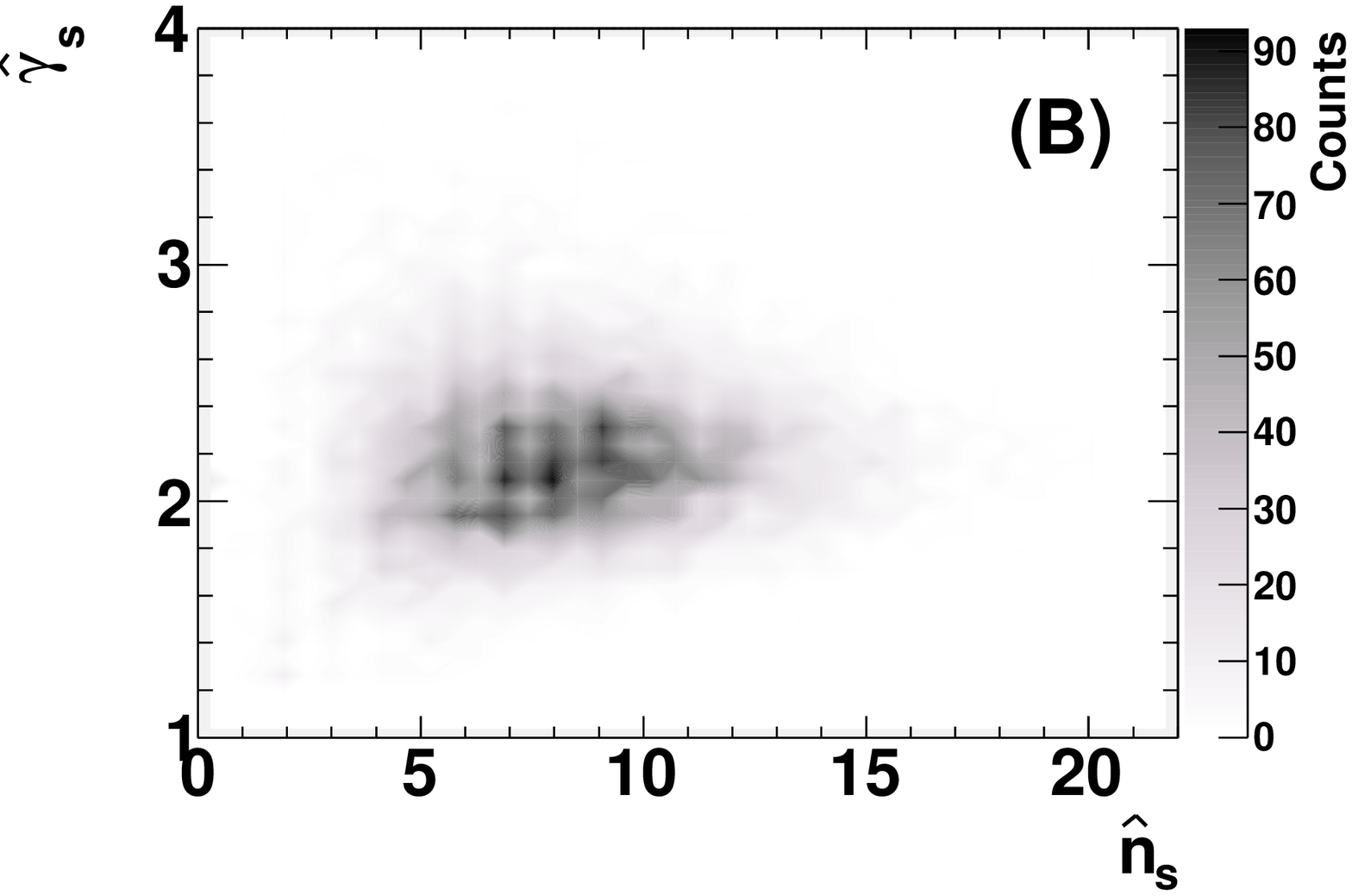}

\includegraphics[width=7 cm,height=5cm]{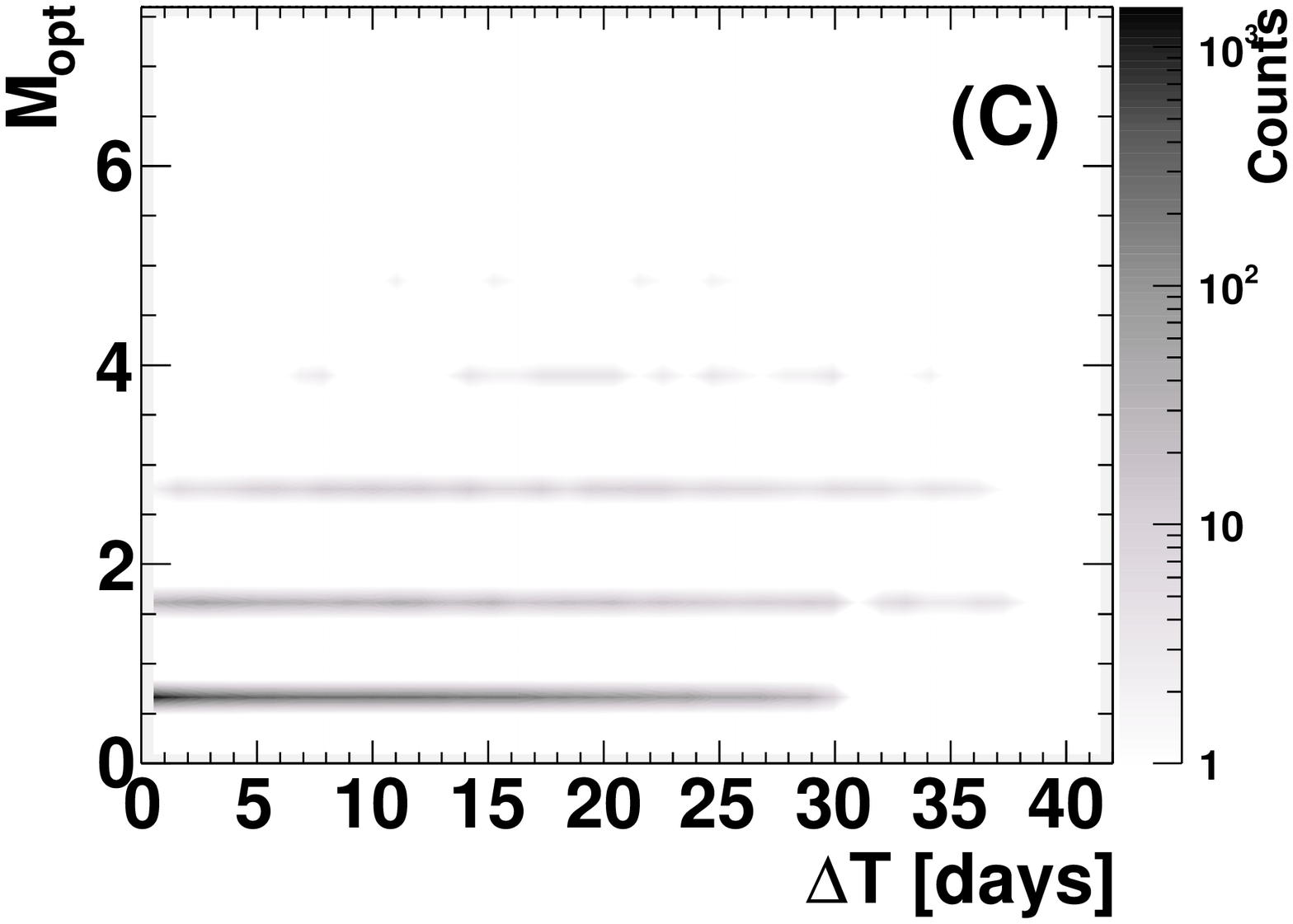}
\includegraphics[width=7 cm,height=5.1cm]{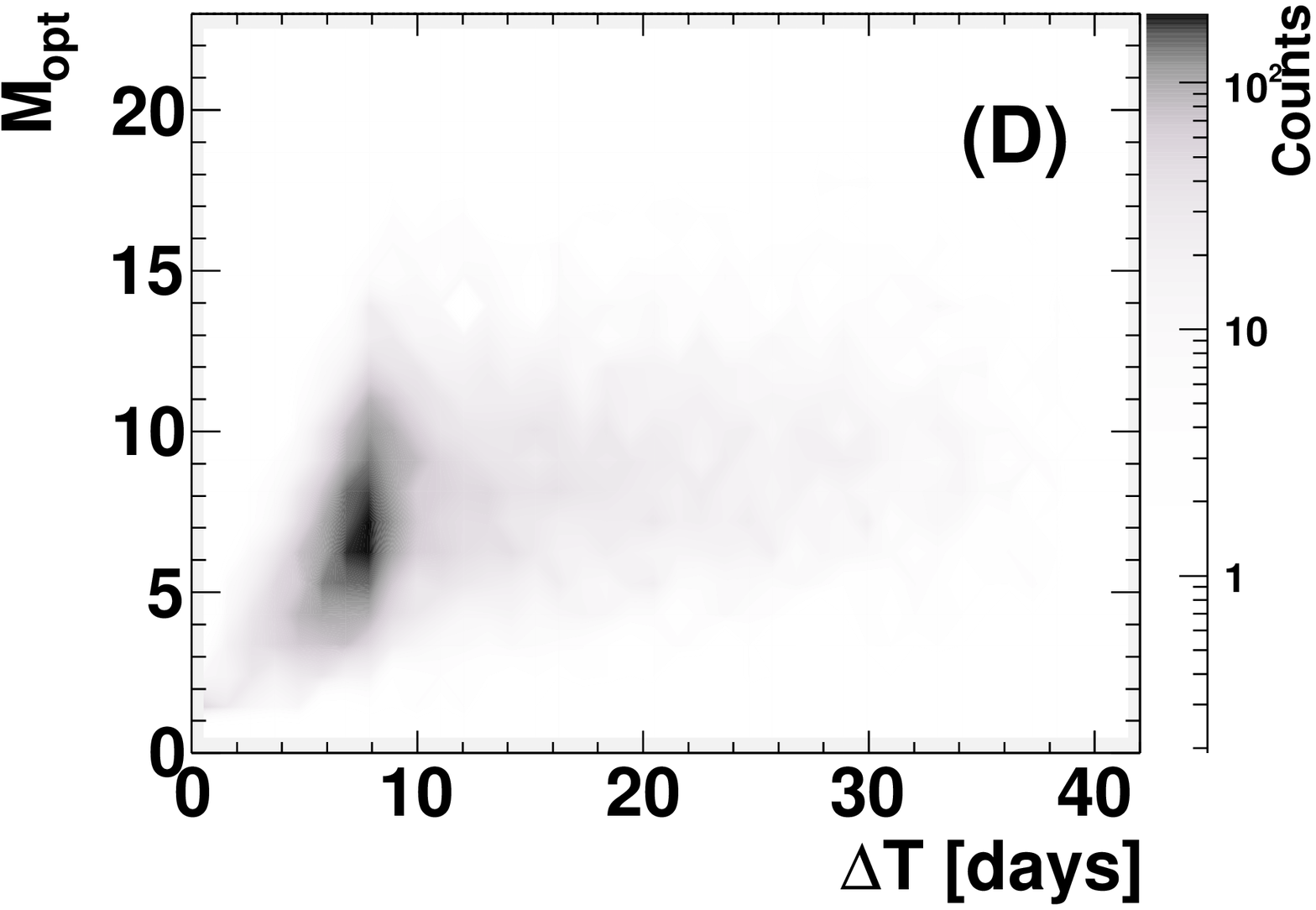}
\caption{\small (A), (B) Distribution of the  best fit values for the number of signal events $\hat n_{s}$ and the source spectral index
$\hat{\gamma}_{s}$   and their correlations 
 for  background-only  simulations (Left)   and   with 8  (Poisson-mean)   signal  events 
 injected  on the  top of the background (Right)  for a  source at declination $15^{\circ}$. (C), (D) The optimal  number of data segments 
$M_{\mathrm{opt}}$ as a function  of the  duration of the flare $\Delta T$ (i.e. the maximum time difference between any data segments  yielding the best fit configuration $M_{\mathrm{opt}}$). }\label{figure4}
 \end{figure*}
\begin{itemize}
\item \textbf{Example~1}: Signal events (with Poisson-mean=8)  are injected  inside
a 9 days time window.   The starting point of the flare is randomly chosen 
over   a data taking period  $\Delta T_{\mathrm{data}}$  of $40$ days.
 An example of a physics case corresponding to this configuration is the  reported $\gamma$-ray flare from \
the Crab Nebula~\cite{agile}, \cite{Fermi}, \cite{argo}.  Note,  that  this case
  would correspond to a   rather  ''strong''  flare. 
As we can see   for example from~\cite{Braun2},
 the number of events   required  for  the  discovery  of such a flare  is larger than 7 in case of  a 
search with  an unknown burst duration.
% {\it  REFORMULATE: The  only requiment is  that  signal events  should be  distributed according to 
%  Gaussian statistics and  should  form  in time  one  single cluster of events (so-called the Gaussian burst).
%  This   is because     standard methods,  which   use  the   one-source  likelihood function  
% given by (Eq.(2)) can  look for one minimum  which corresponds only to  the most significant
%  cluster of  events  from a given source location.  }

\item \textbf{Example~2}: The  total number of signal events (Poisson-mean=8)
 is  the same  as in  Example~1, but individual  events  are injected  over  two  time windows 
of duration 4.5 and 9 days, respectively and 22 days apart.
The   average number  of  injected signal events   is  the same  for each flare (Poisson-mean=4).
The expected  number of events  required for  the discovery 
depends  on the flare duration and it is     smaller for  the  flare
 with  shortest duration. 
In  other words we can think  of  this  example   as a case with  a  ``strong''  flare 
and  one  weaker flare. These two   flares  are separated in time, but   
they   can also  be  treated as one long period of enhanced emission  with  a  duration  about of 35 days.   Standard point  source search  methods applied  to this case
 will find only the strongest flare i.e. the first one. This   is because     standard methods 
  use  the   one-source  likelihood function   given by  Eq. (\ref{llh_unbinned})  and can  look  only for a minimum  corresponding  to  the most significant cluster of  events  from a given source location.

\item \textbf{Example~3}:  The total number of signal events  (Poisson mean=8)
is the same as in previous examples, but individual events are injected over    three time windows with duration of 4.5 days, 4.5 days and 9 days, respectively. 
  Each flare is simulated with a  similar  strength with Poisson-mean 3, 3 and 2, respectively.
 This example  describes  three  ``weak'' flares. Note that the  number of events  required to claim discovery for  each of them 
 at  a $5 \sigma$  level is larger than 4, as we can see from \cite{Braun2}~\footnote{In this work we 
injected signal events according to a uniform distribution while in \cite{Braun2} a Gaussian distribution was considered. While a general comparison is still possible in some cases  
 the exact numerical results are different. The duration of a time window, for example, is somewhat different. For  a  uniform distribution   with non zero  values in the interval $\Delta t$   the standard deviation is defined  according  to $\sigma=\Delta t /\sqrt{12}$ with  $\sigma$    denoting  the standard deviation of the Gaussian distribution. Thus
 4.5 and  9  days width of the flare  corresponds to  $\sigma_{t}\simeq \sigma$ about  1.3 and  2.6  days or
 to 0.0035 and 0.0070  year, respectively. Thus for  the flare with a Gaussian mean time $\sigma_{t}=0.0035$ we need  about 4 events for  discovery using the method labeled by ``Assumed Burst Time (Energy)''  in Figure 4 of \cite{Braun2}.}.  
In   other words such  individual weak flares  cannot be  found at a $5\sigma$ level by  the  standard  point sources search  methods.
\end{itemize}

\subsection{Example 1: one flare}
 
In Figure~\ref{figure4}  the distribution of the most important parameters i.e.  the 
 number of signal events $\hat{n}_s$,  the spectral index $\hat{\gamma}_{s}$  and 
the optimal number of data segments (i.e. signal-like doublets)  $ M_{\mathrm{opt}}$ 
 and their correlations  as obtained by MC simulations  are presented for  the  background-only  simulations  and   the   background plus  signal  cases.  
The number of  injected signal  events   follows  a Poisson distribution with  a mean of 8.  Events are  injected  over a time window   of  9 days  duration.  
The pure  background case  illustrates  how signal-like flares can be mimicked   by pairs  of events (sometimes also  triplets) 
distributed  over very  short time  windows with durations of  less than  a day. The signal plus background plots show   that the 
 proposed method finds  the right  flare i.e. it  recovers the true values of the  spectral   index (2),  the 
 number of injected signal events (Poisson-mean 8 events)  and the    flare duration~\footnote{The exact duration of the flare  can be slightly less since the signal events were randomly injected inside the 9 days time window.}  (9 days). In addition,    the
  flare  can be  decomposed  into  about  8  data segments, see Figure~\ref{figure4} (D).
\begin{figure*}[ht]
 \centering
 \includegraphics[width=7 cm,height=5cm]{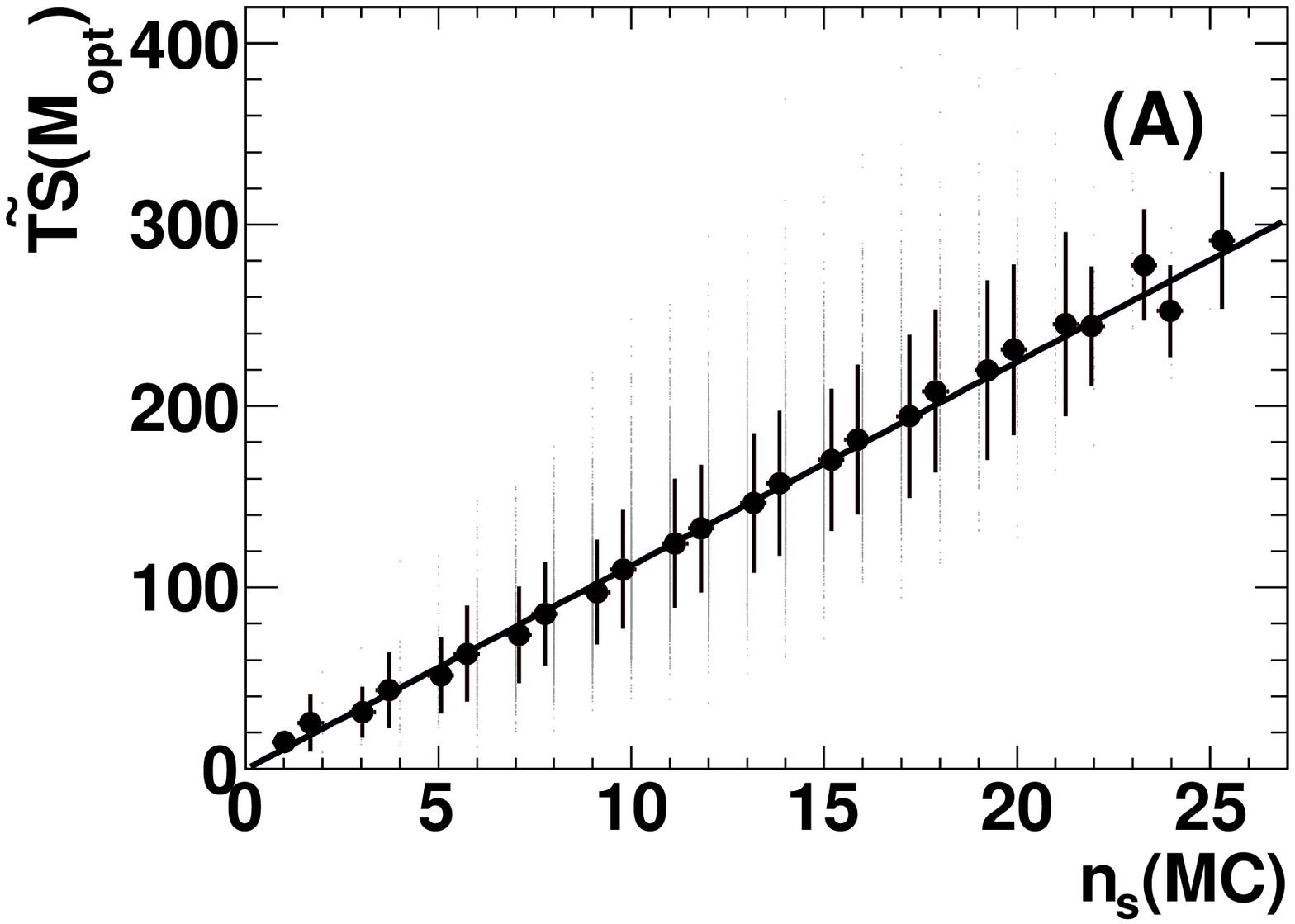}
 \includegraphics[width=7 cm,height=5cm]{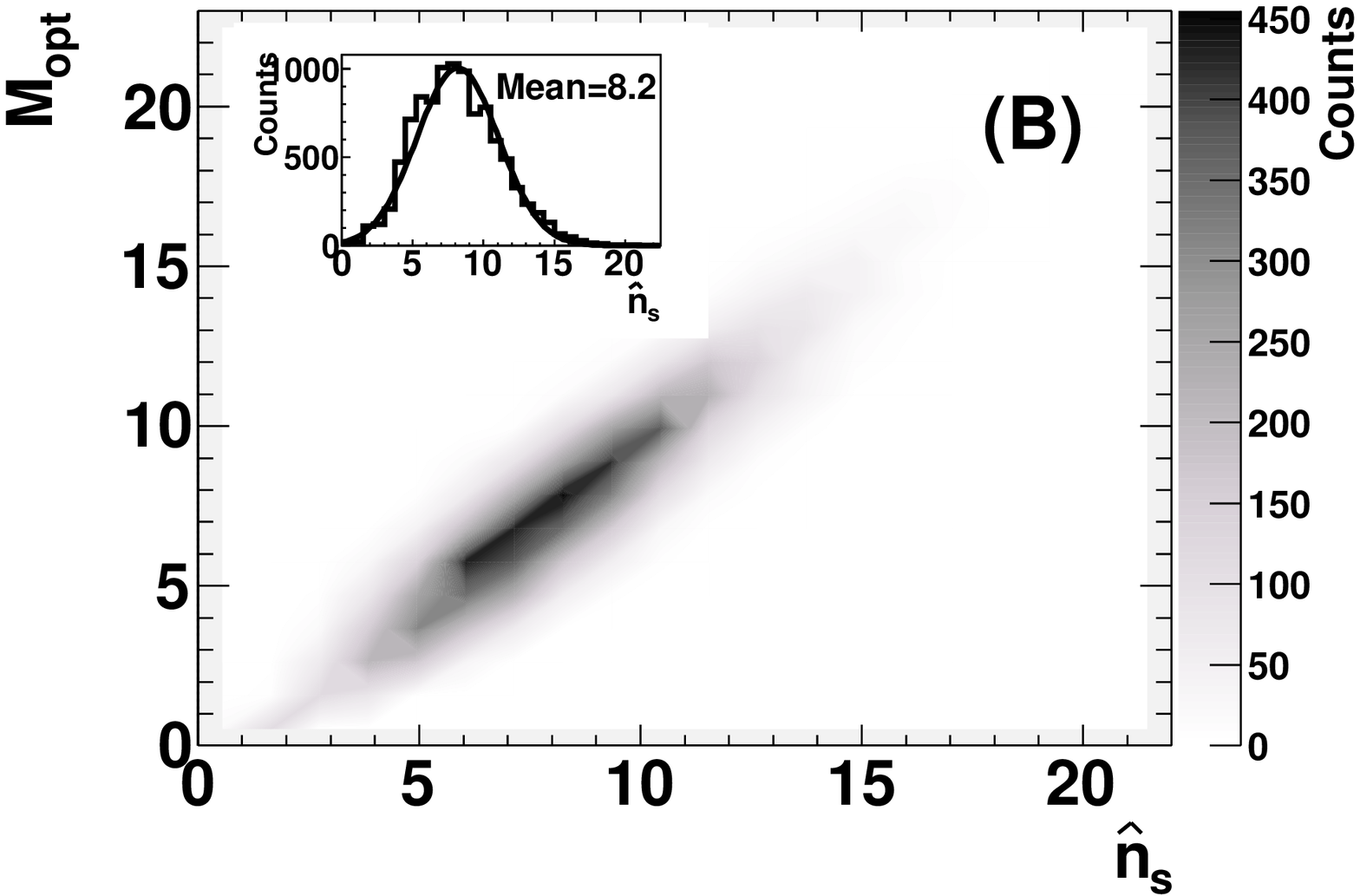} 

 \caption{\small (A) Correlation of  the test statistic and  
 the true number of  injected signal events.  The solid line represents  a linear  fit to  data.  Full circles  correspond  to  an average value  
of the test statistic for  a fixed number of injected signal events. 
 (B) The  optimal number of doublets  which compose the flare with  9 days duration
  as  a function of   the average number of expected signal events.
 The  inset  shows  the distribution  of the best fit values of the number of  signal  events which can be well fitted with a   Gaussian.}\label{figure5} 
 \end{figure*}

In Figure~\ref{figure5} (A) the correlation  of the global test statistic  $\widetilde{\mathrm{TS}}(M_{\mathrm{opt}})$  as a function  
of  the number of injected signal events $n_{s}(\mathrm{MC})$ is shown.  The solid
line represents a linear fit   to the  mean  values of the test statistic  calculated for  a fixed number 
of  injected signal events.  A linear  correlation is found  between  the test statistic and  number of injected signal events.
This justifies using  the test statistic  as a weight in  the modified likelihood $\tilde \mathcal{L}$ of  Eq. (9).  
We also checked the results obtained    when  all weights are  fixed to  one 
as well as  to the square of  the test statistic. We find that such  
modifications lead  to  slightly worse results i.e. we  observed   a  5-10\% worse  
agreement between  the  true  and the  estimated values  of  $\hat{n}_{s}$ and  $\hat\gamma_{s}$.

In Figure~\ref{figure5} (B)   the   number of  signal events  is shown   as a function of the optimal number ($M_{\mathrm{opt}}$) of
data segments  which compose the flare. 
 As  expected, the  number of  signal events increases when   more  segments are being added. 
The distribution shows a maximum  corresponding to  the  true  number 
 of injected signal events (the true Poisson-mean  of 8).  The  inset in   Figure~\ref{figure5} (B)  shows 
 the  Gaussian fit to the  estimated number  of signal  events, $\hat{n}_s$  with  a mean value of 8.2,
which is  good agreement  with the  true value  at  a level  of about  2.5\%. 

The  proposed algorithm can  therefore recover the most important parameters  characterizing  flares,
like   the  duration, the  source spectral index and  the number of injected
 signal events with  uncertainties not larger than  a  few percent. In  other words  the proposed 
algorithm can ``decompose``'  the flare  into small data segments which contain signal-like doublets 
and  can   effectively   reject   
 background events  from  the  entire   data period. 
\begin{figure*}[ht]
 \centering
 \includegraphics[width=0.45\textwidth,height=4.1cm]{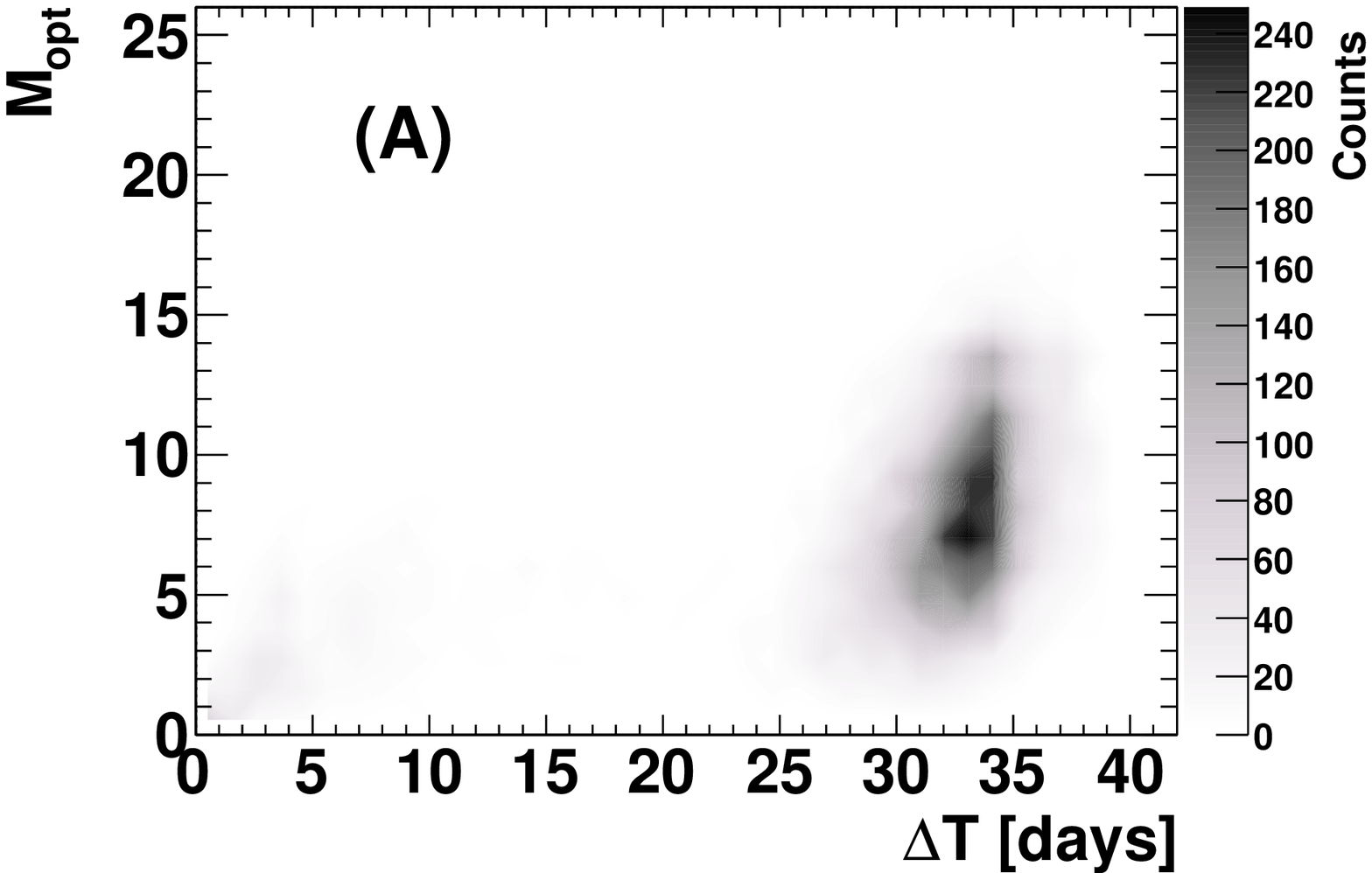}
 \includegraphics[width=0.45\textwidth,height=4.1cm]{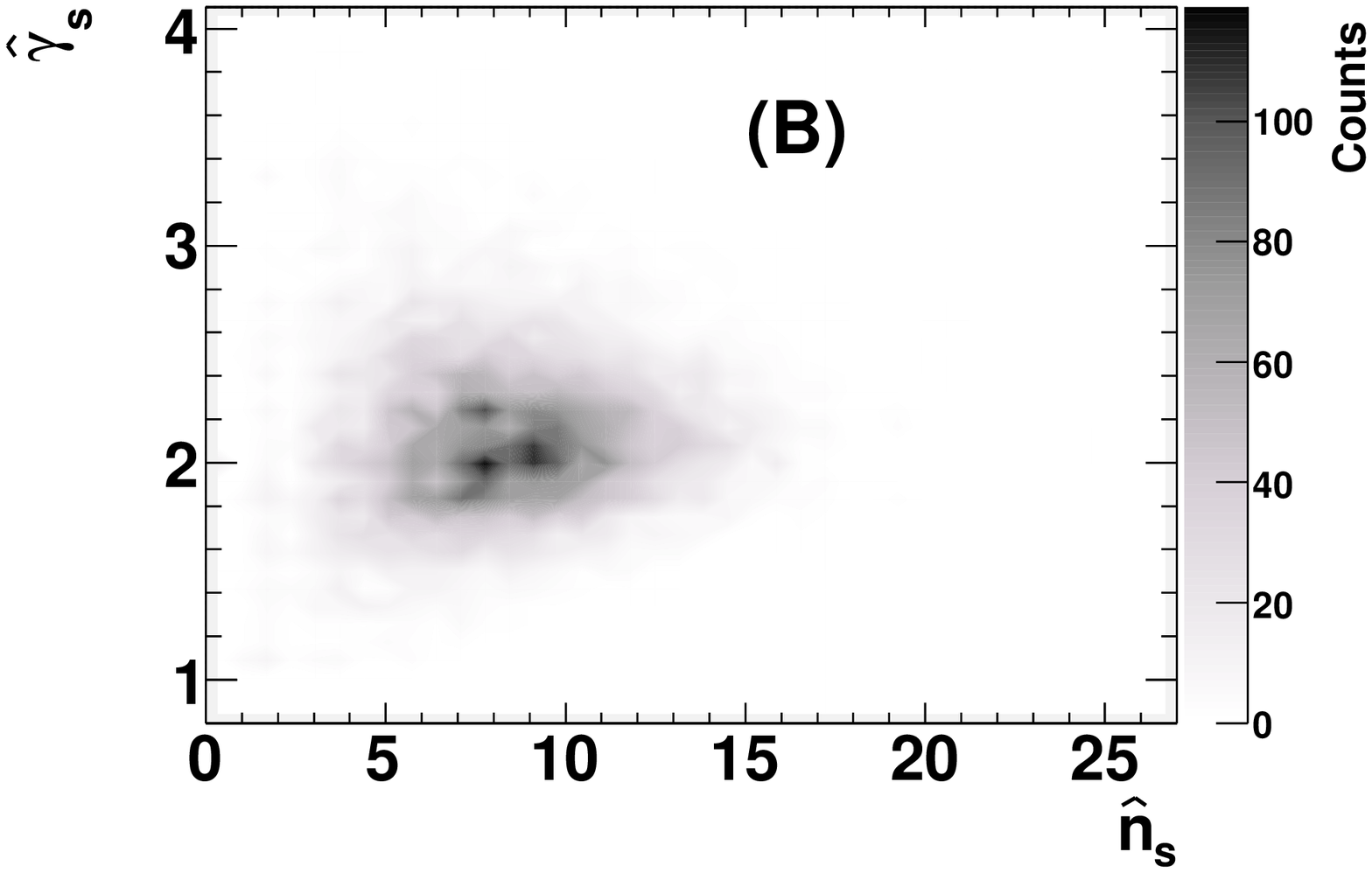}

 \includegraphics[width=0.45\textwidth,height=4.1cm]{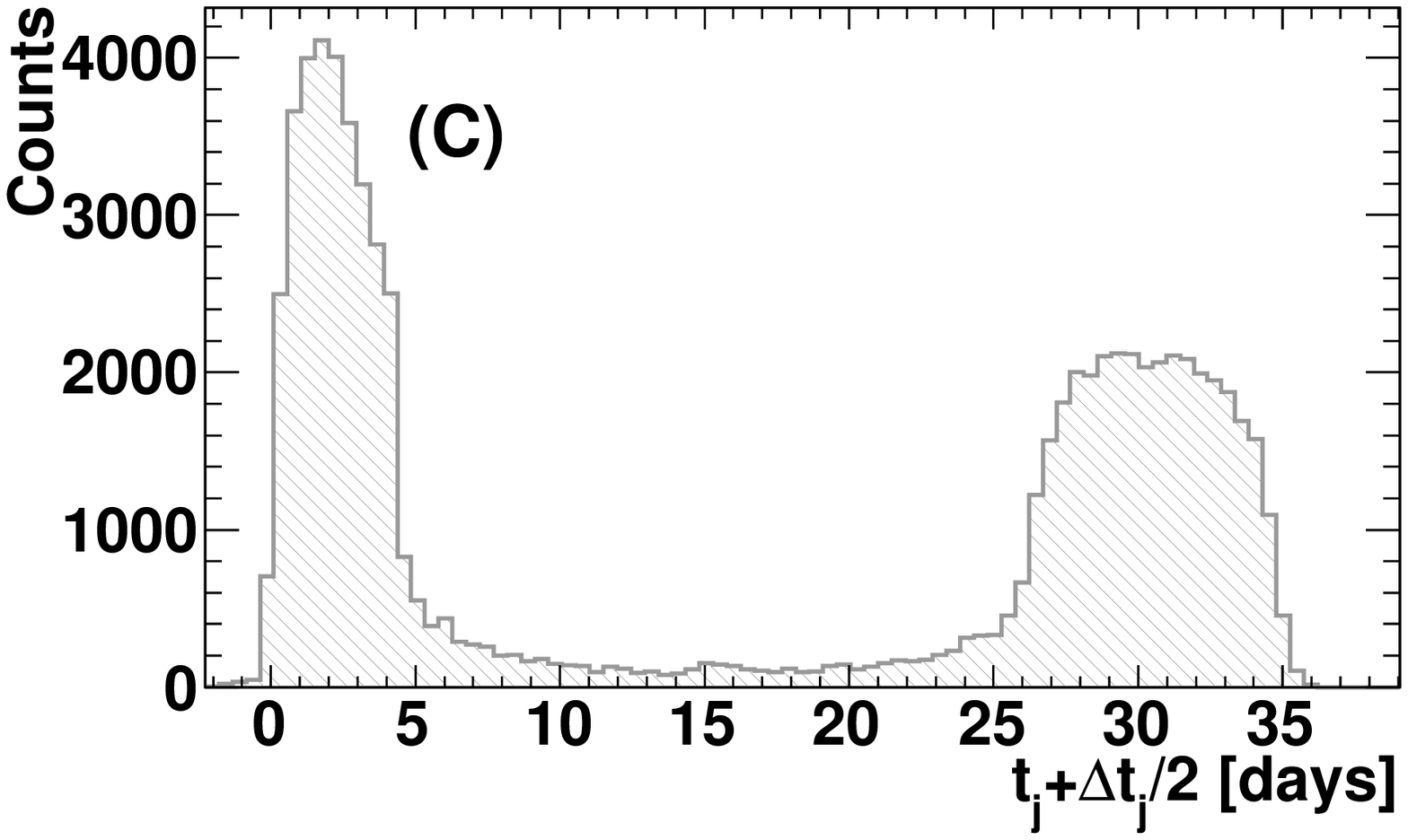}
 \includegraphics[width=0.45\textwidth,height=4.1cm]{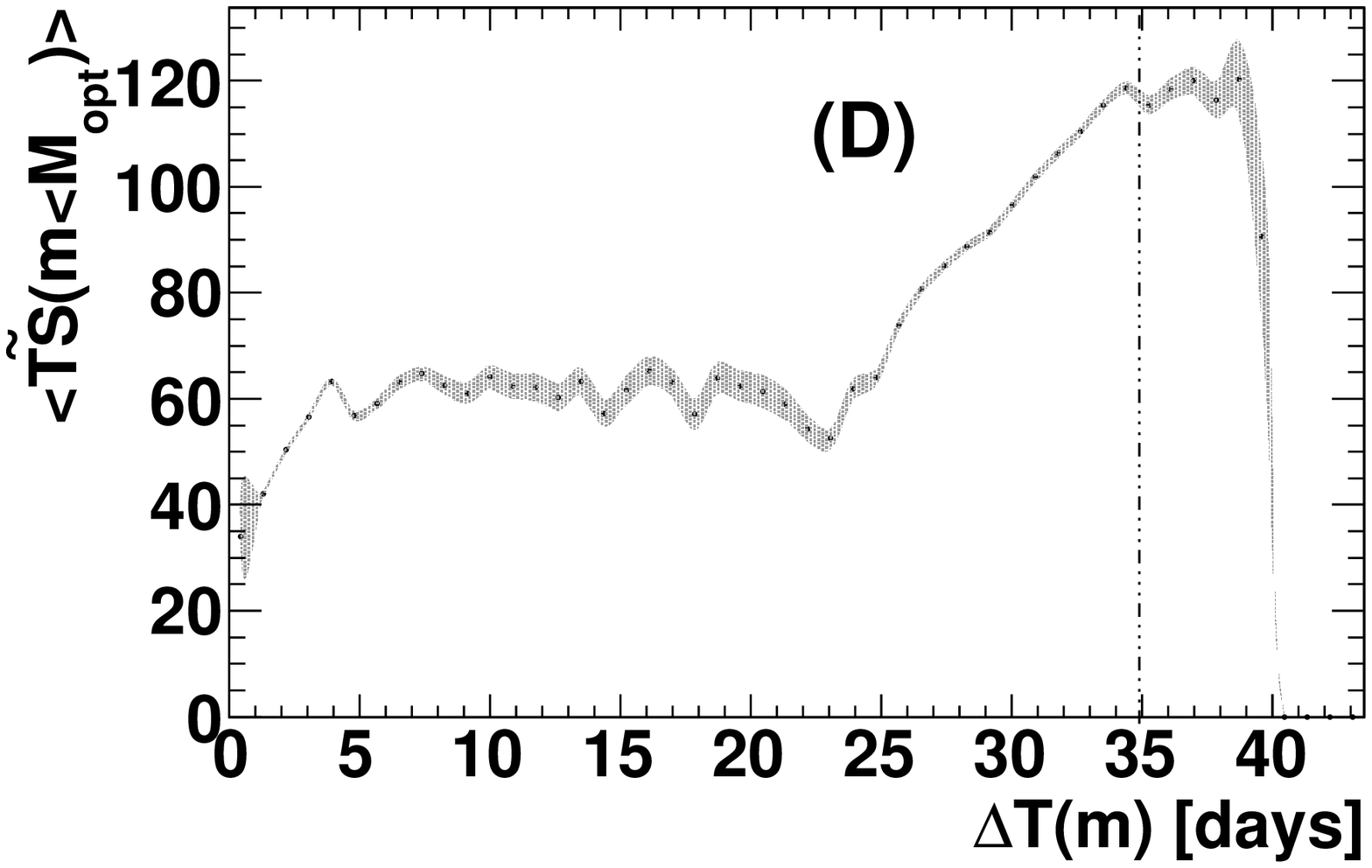}

 \includegraphics[width=0.45\textwidth,height=4.1cm]{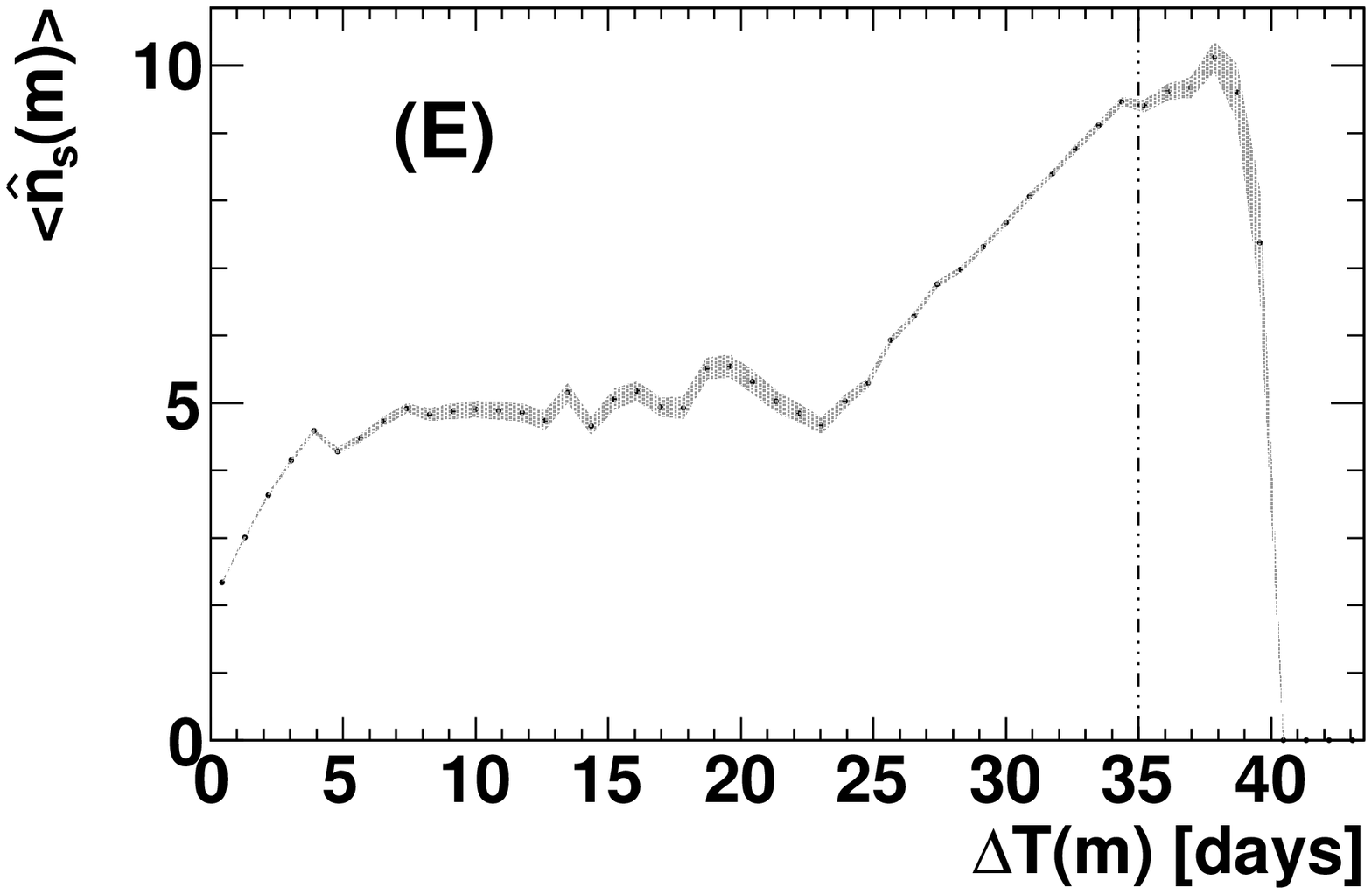}
 \includegraphics[width=0.45\textwidth,height=4.1cm]{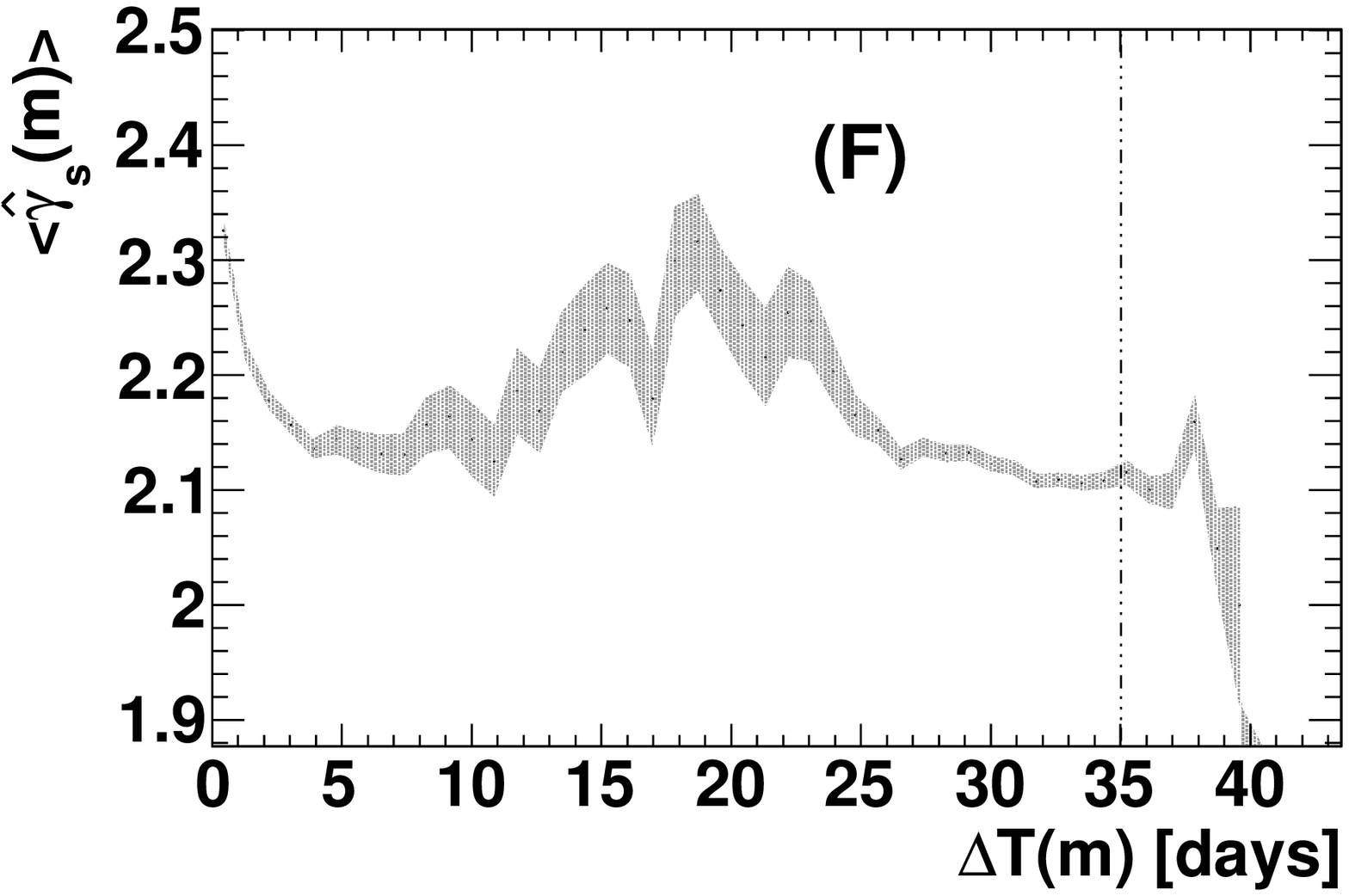}
 \caption{\small Example~2: Results of the search  for two  individual flares separated  in time
 using the method presented in this work. (A) Distribution of the  optimal number of doublets
($M_{\mathrm{opt}}$)  as a  function of the total flare duration $\Delta T$; (B) The spectral 
index versus the  expected number of signal events; 
 (C) The  mean  time of each data segment; (D), (E), and  (F) 
The  distribution of $\left< \widetilde{\mathrm{TS}}(m) \right>$, $\left<\hat{n}_{s}(m)\right>$ and spectral index $\left<\hat{ \gamma}_s(m)\right>$ as a  function of the  flare duration $\Delta T(m)$. Also shown are 1-$\sigma$ ranges.
The vertical  dashed line indicates the overall period of enhanced  emission (35 days). See the  text for more details.}\label{figure6} 
 \end{figure*}

 \subsection{Example 2:  two flares separated in time}

In Figure~\ref{figure6}~(A)  we show  the   distribution of the  optimal number 
of  data segments  $M_{\mathrm{opt}}$ as  a function of   the  corresponding  
flare  duration $\Delta T$ as defined in section 2.3.
In Figure~\ref{figure6}~(B)  the  best fit  spectral index  $\hat{\gamma}_{s}$   is shown as a function of the  estimated number of signal 
events $\hat{n}_s$.  Also in this case  the  proposed algorithm 
recovers  the overall  flare duration, the  source  spectral index and  the number of injected signal events. However, for this particular case, 
the  differences between the true  and the  estimated  values of the  physics  parameters obtained  from the minimization are  slightly  worse  (about  $5\%$). This is  because  for the  larger flare durations we should  expect a larger contamination   of background events.  

In Figure~\ref{figure6} (C) the distribution of the mean time  $T_{0,j}=t_{j}+\Delta t_{j}/2$  calculated for each  ($j^{\mathrm{th}}$)   signal-like 
 doublet    is shown.  An  accumulation of signal-like  doublets  for a time  period below 
5 days   and   between 25 and 35 days  is visible, corresponding to the first and the second flare, respectively. 
The  proposed algorithm can therefore find not only the most significant flare,
but also the  weaker  one  separated in time. Such a flare  would not be detected by other existing  methods (like~\cite{Braun2}).
\begin{figure*}[ht]
 \centering
 \includegraphics[width=0.45\textwidth,height=4.1cm]{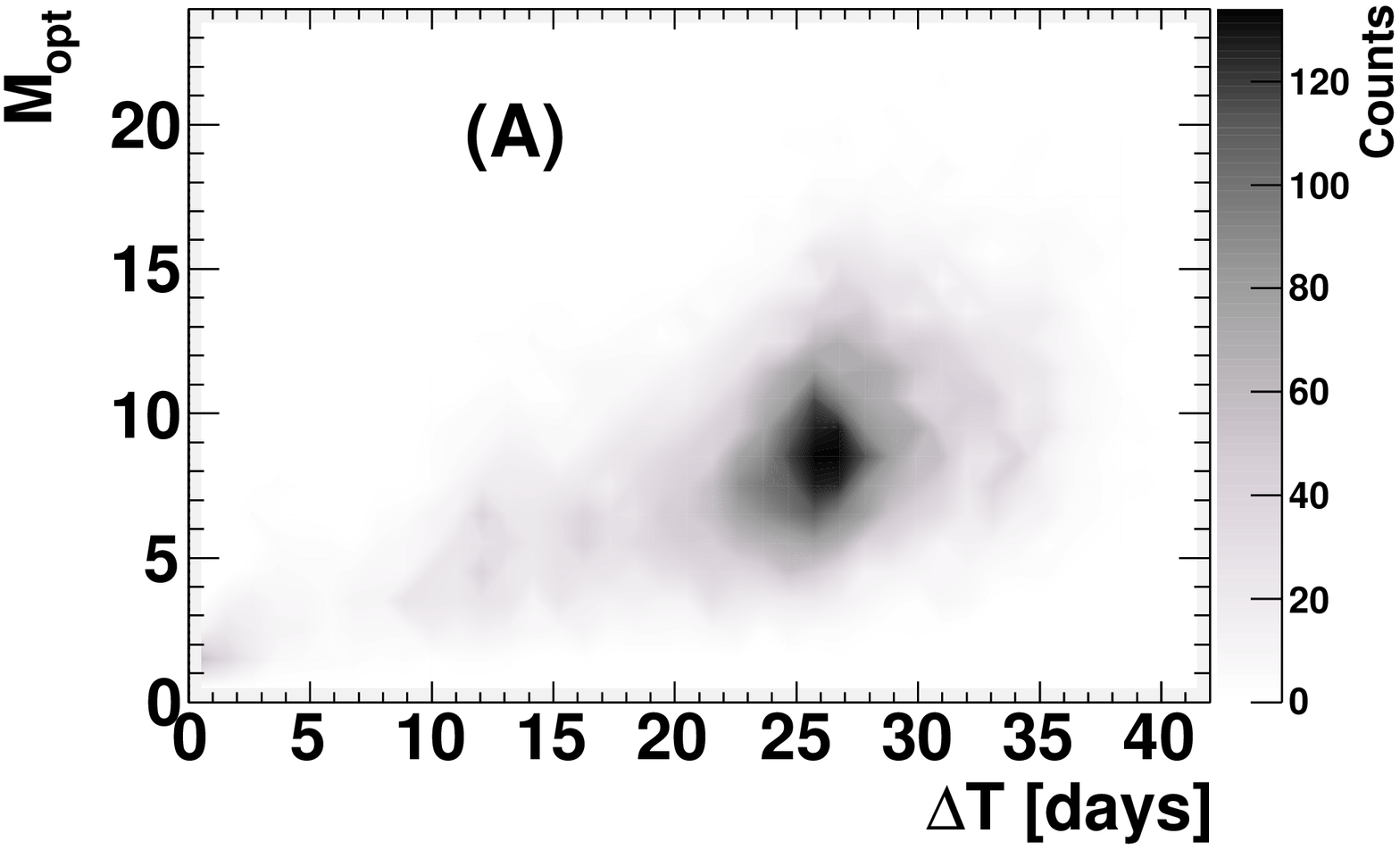}
 \includegraphics[width=0.45\textwidth,height=4.1cm]{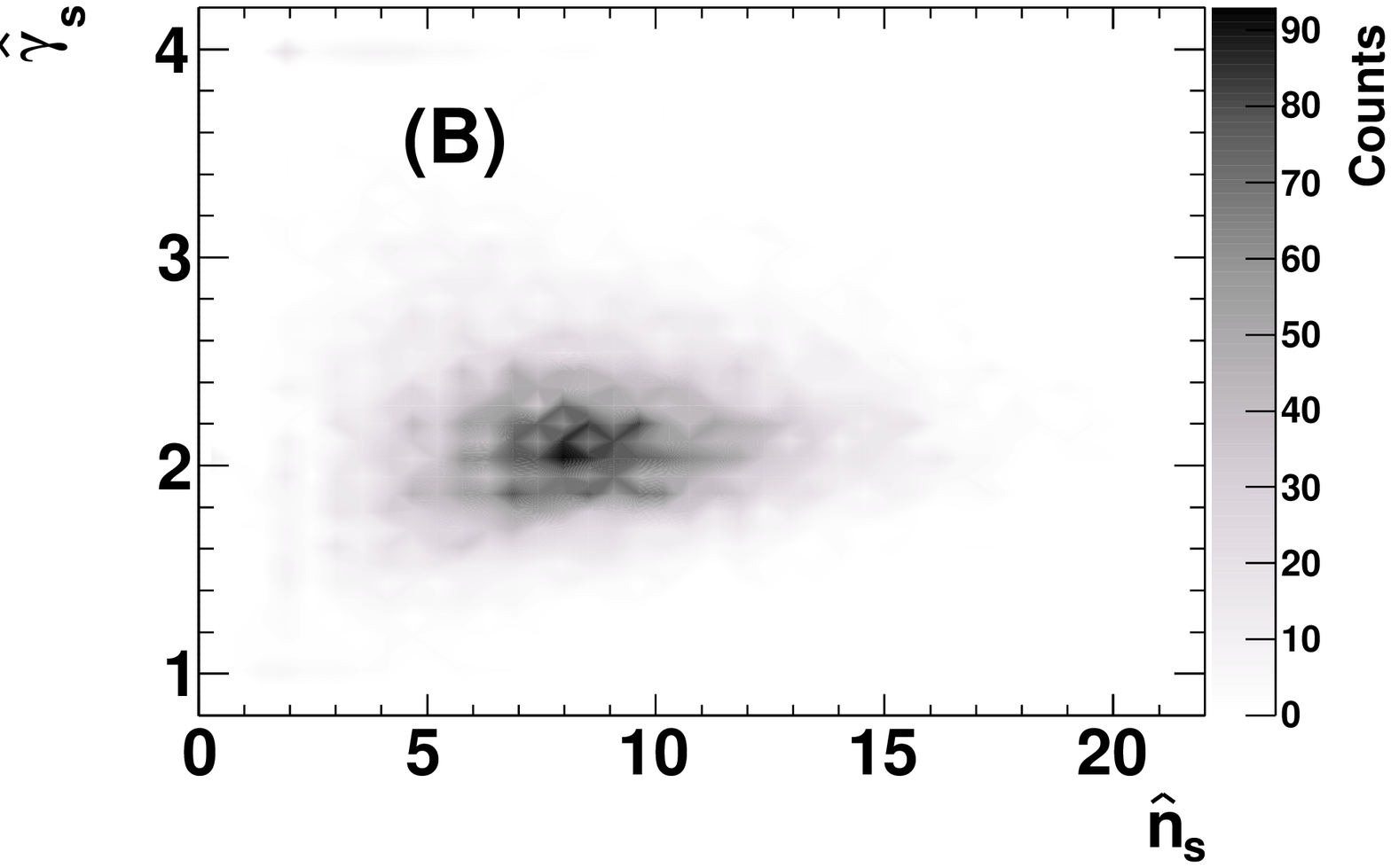}

 \includegraphics[width=0.45\textwidth,height=4.1cm]{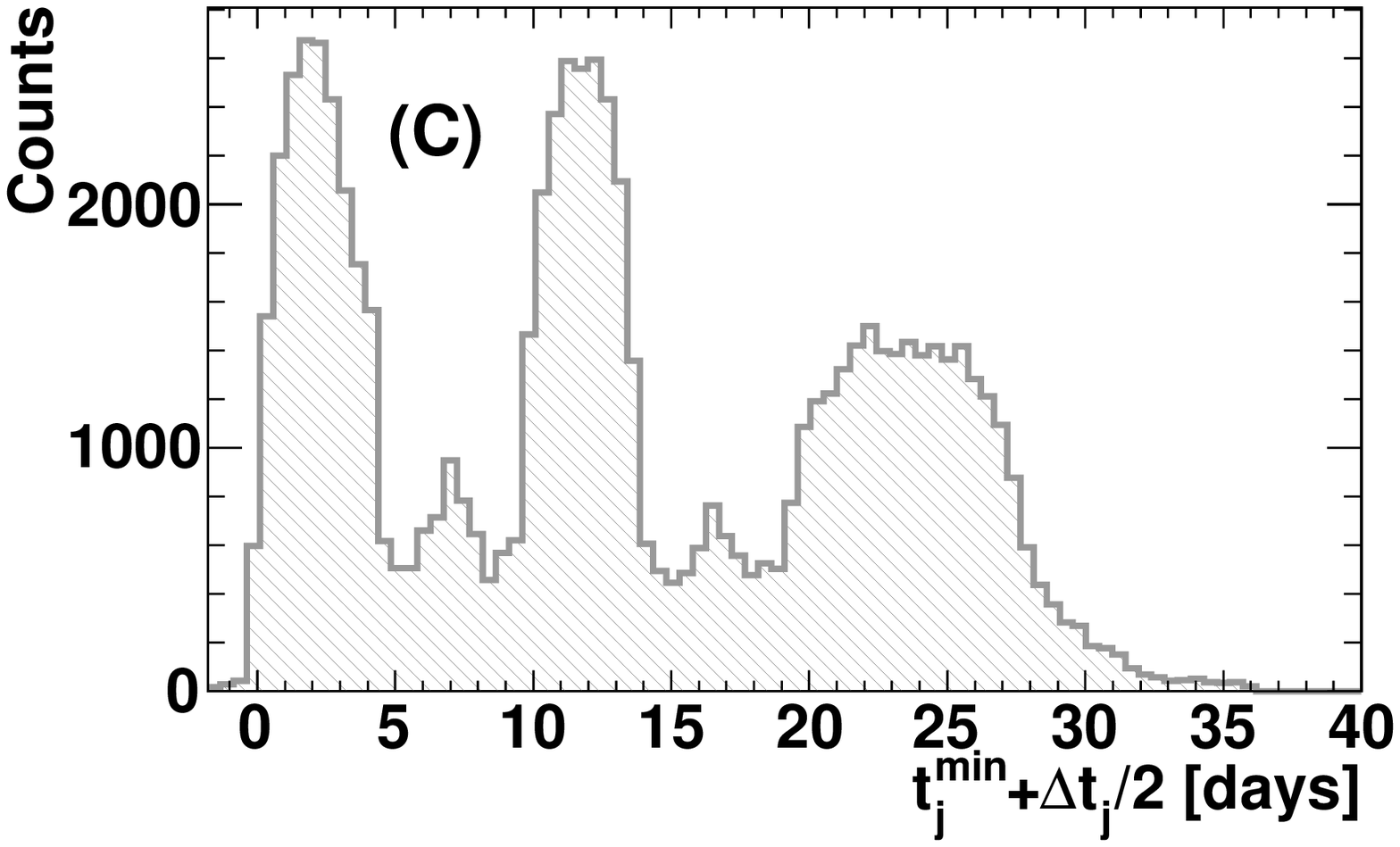}
 \includegraphics[width=0.45\textwidth,height=4.1cm]{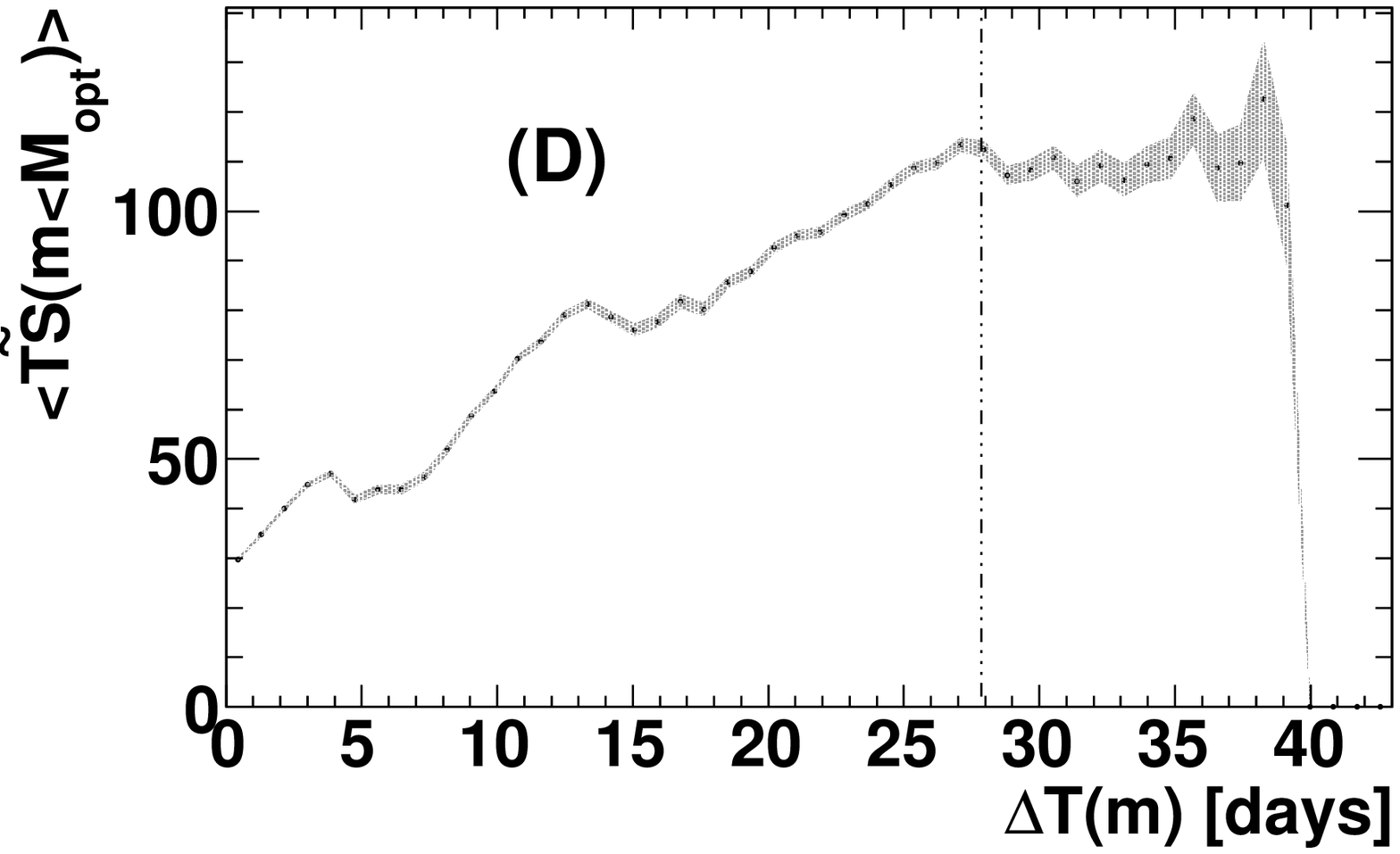}

 \includegraphics[width=0.45\textwidth,height=4.1cm]{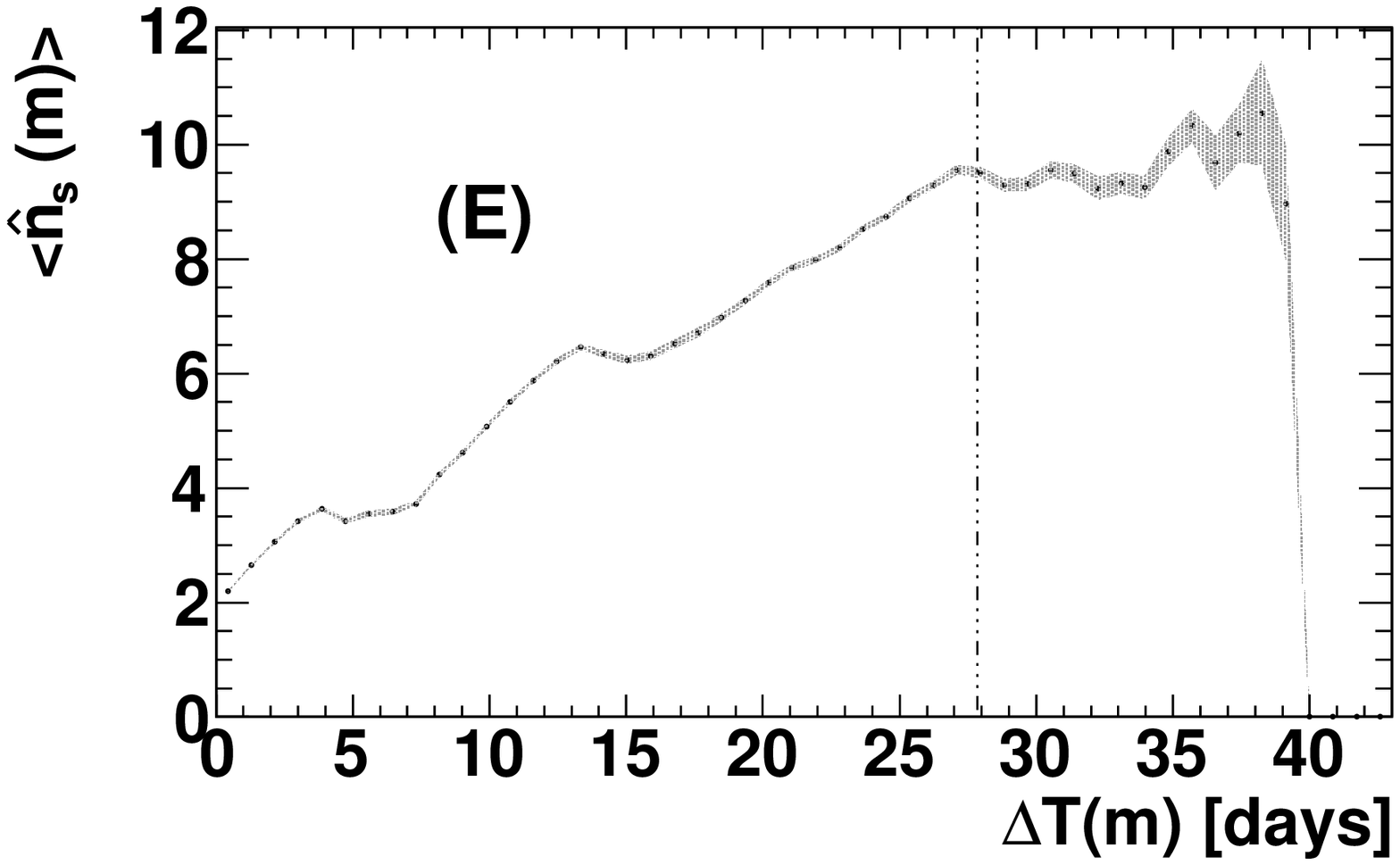}
 \includegraphics[width=0.45\textwidth,height=4.1cm]{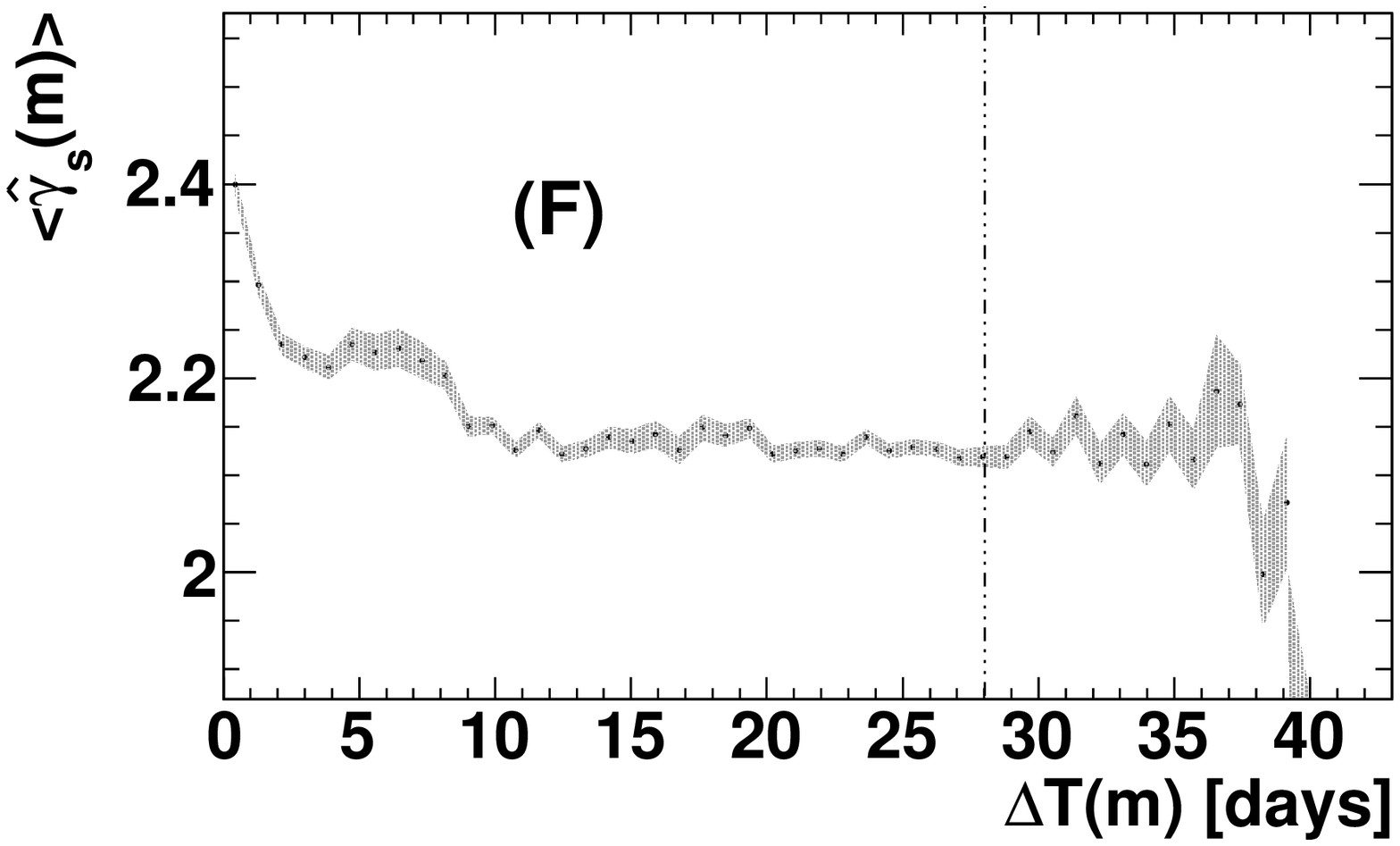}

 \caption{\small Example~3: Results of the search  for three  individual flares separated  in time.
 (A) Distribution of the  optimal number of data segments
($M_{\mathrm{opt}}$)  as  a function of the total flare duration $\Delta T$; (B) The spectral 
index versus the expected number of signal events; 
 (C) The  mean time for   each data segments;
(D), (E), and (F) The distribution of $\left< \widetilde{\mathrm{TS}}(m) \right>$, $\left< \hat{n}_{s}(m)\right>$ and  the spectral index $\left< \hat{\gamma}_s (m)\right>$, respectively. The vertical  dashed line indicates the  overall period of enhanced emission   (28 days). See the text for more details. }\label{figure7}
 \end{figure*}

By repeating the minimization with the  modified likelihood  $\tilde \mathcal{L}$ for  a fixed
number of $m$  data segments  {\it sorted in time} (starting from $m=1$ and 
ending with $m=M_{\mathrm{opt}}$),  we can get  more information about the ''internal''  structure of the flare. 
% Note,  that now  index $m$ counts data segments  sorted according to time  as it shown in Figure~\ref{figure3} (Top))  but not like   before  according to the value of the  individual test statistic $\mathrm{TS}_{j}|_{\Delta t_j}$, Figure~\ref{figure3} (Middle). 
For example, in Figure~\ref{figure6} (D)  changes of the  global  test statistic as a function  of   the flare duration  $\Delta T(m)$~\footnote{
For a   fixed number $m$ of data segments,
the flare duration is   defined  as the time between the start  time of the first  in time  data segment  
and the end time of the last data segment. } calculated for $m<M_{\mathrm{opt}}$ data segments  is shown.
The  test statistic increases  when more  segments  are added  in time, but 
 finally reaches a saturation   for the   time corresponding to the  overall period of enhancement emission  (35 days). This behavior
 is  strongly correlated with the distribution  presented in Figure~\ref{figure6}~(C). For  the first flare (below 5 days) 
and second flare (from 25 to 35 days) the increase  of  the global test statistic in time   is  
 much larger than  during the   period    corresponding  to  the  22 days time gap. During this last period larger variations can  also be observed, 
which is an indication for background fluctuations. A similar  behavior is  
  seen  for  the  average number of signal events  $\left<\hat{n}_{s}(m)\right>$   
 as  a function of the  flare duration (Figure~\ref{figure6}~(E)). Note that  
both  distributions presented  in Figure~\ref{figure6}~(D) and (E)  are  averaged  
over  different MC realizations.  

For completeness,   Figure~\ref{figure6}~(F)  shows  the average spectral index $\left<\hat\gamma_s(m)\right>$    for a large number of  different
 MC realizations as  a function of time. The average  best fit  spectral index  is about 2.2 and  differs 
 by  about  8.5\% from the true value ($\gamma_{s}=2$).  Similarly to the previous distributions, the  
fluctuations  of the  spectral index  are strongly reduced during the  periods
 corresponding to the injected signal. The  effect is particularly visible for the second flare, between  25 days  up to  35 days. 

We conclude  that the   proposed algorithm  can  recover  the  global parameters describing the flares
and provide additional   information about  the  time  development, their  duration
 and    the  number  of  signal events per individual flare, even when they alone  are  below 
the threshold for detection.

\begin{figure*}[ht]
 \centering
 \includegraphics[width=0.48\textwidth,height=4.5cm]{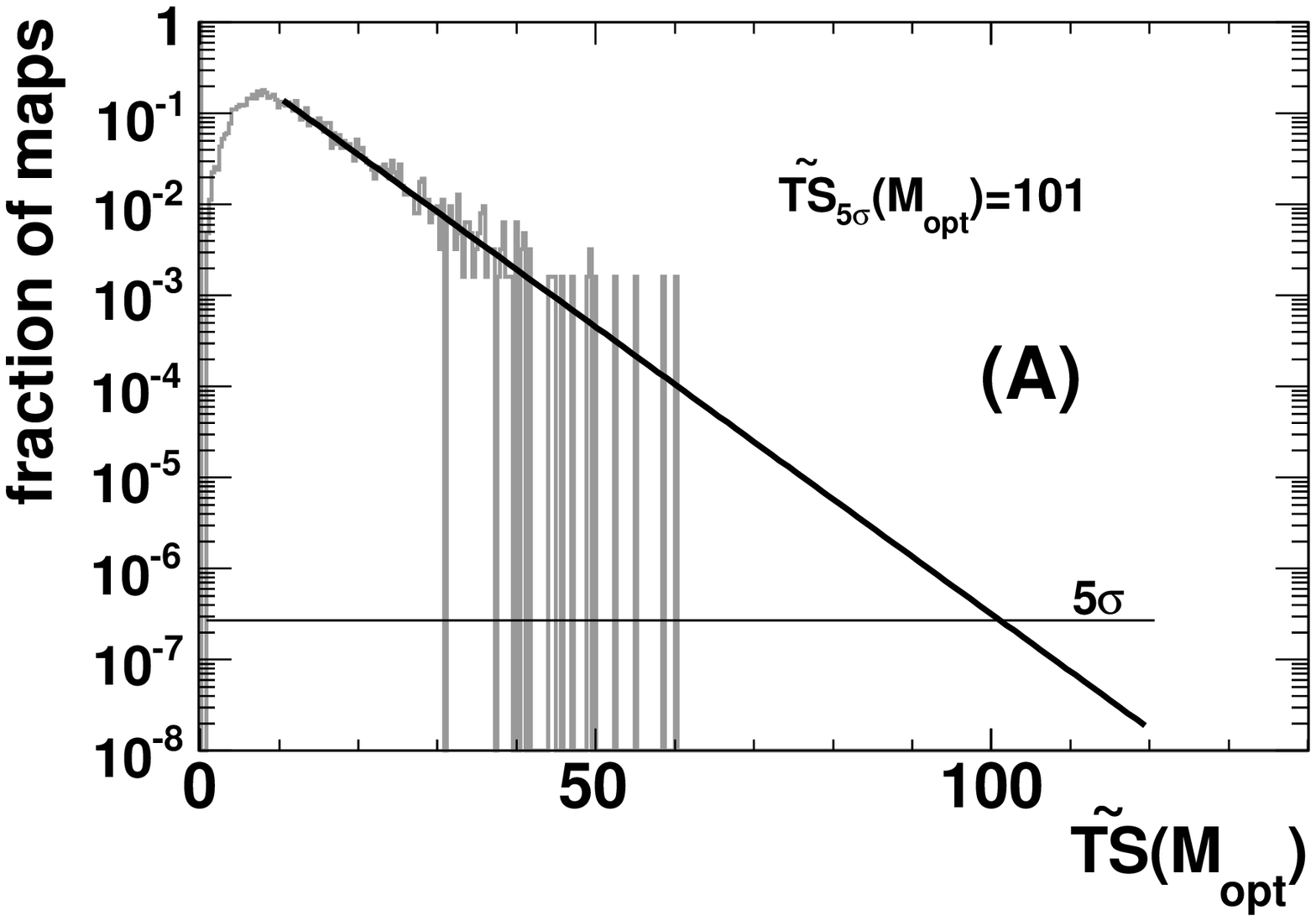}
\includegraphics[width=0.48\textwidth,height=4.5cm]{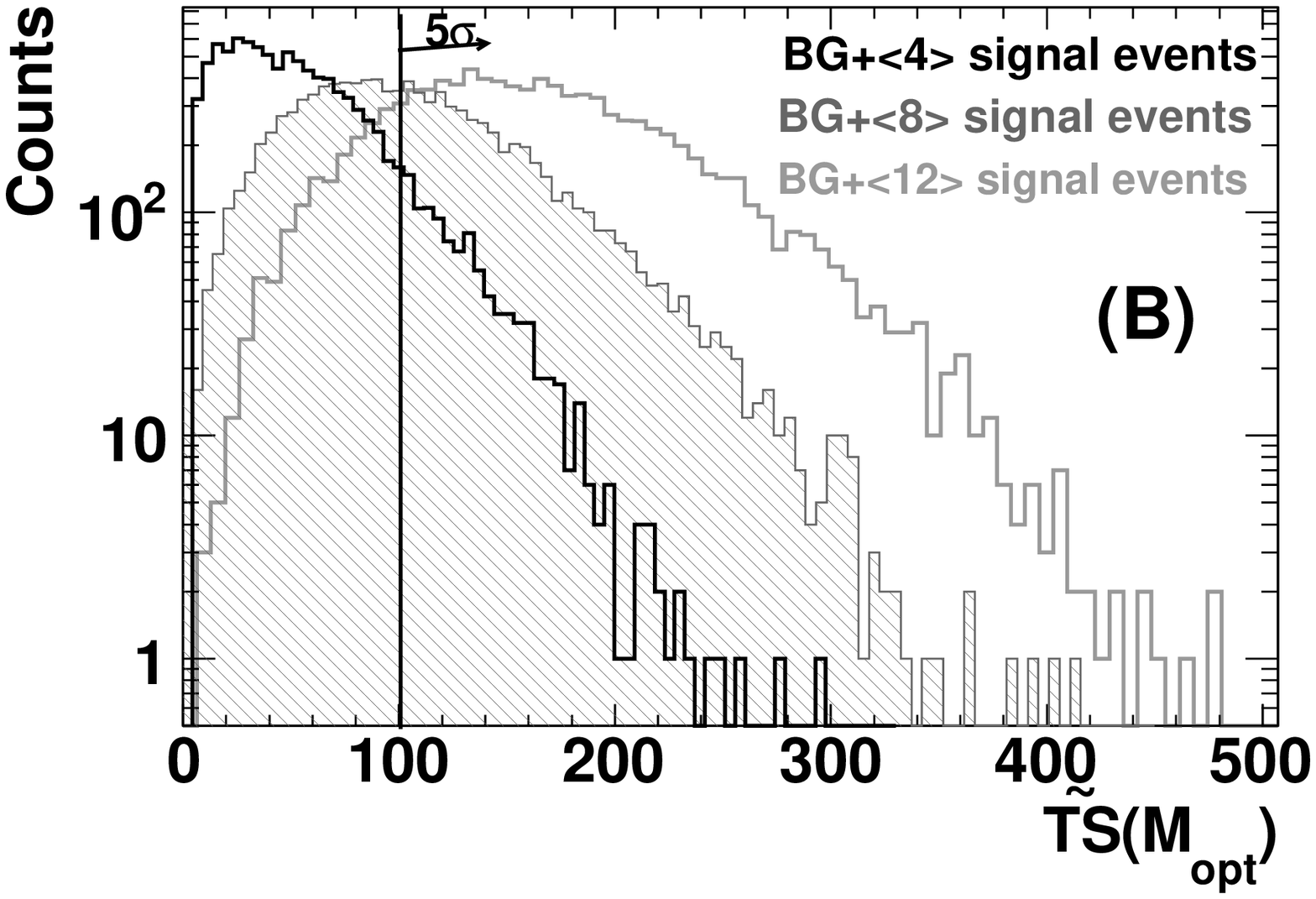}
 \caption{\small (A) Probability distribution  of the global test statistics $\widetilde{\mathrm{TS}}(M_{\mathrm{opt}})$ 
for  background-only simulations. The  solid  line  shows an exponential fit   to  the distribution
used to obtain the  $5\sigma$  threshold for the background only case; (B) Distributions of 
the  test statistic $\widetilde{\mathrm{TS}}(M_{\mathrm{opt}})$ for  background plus  signal simulations, where signal events
are injected  according to  a Poisson distribution with  mean:  4, 8, 12  signal events, respectively. The  vertical line corresponds to the $5 \sigma$  threshold.}\label{figure8}
 \end{figure*}

\subsection{Example 3:  three weak flares separated in time}

In this case  we simulated three  individual weak  flares.  
Each flare cannot be individually detected  at  a $5\sigma$ level by   using  the  standard point source search  methods.
In principle, if all events could be injected into a single   flare of  duration about 28 days ($\sigma_{t}\simeq0.02$ year)  
 the algorithm proposed in~\cite{Braun2} would yield  a discovery. For a flare of  such a  duration,  
more than about  7  events  are needed   for  discovery at  a $5 \sigma$ level  using  the 
 method labeled by Assumed Burst Time (Energy) in Figure~4 of~\cite{Braun2}. However, the flare 
will be found  only   if  these 7 events   will   form {\it one  cluster of events compact in time}. This is because
the unbinned  maximum likelihood method  with a single-source likelihood function can only search  for a  maximum 
which corresponds to  the  most significant flare (cluster of events) from  a point  source. 
This is  certainly  not the case for this example, where  8 events  are  distributed  
among three individual and well  separated  flares.  Thus, such a structure   
cannot be detected at $5\sigma$ level by the standard method.
%~\footnote{In addition,  for Example 3
%we performed a  time-integrated  analysis ($P^{\mathrm{time}}=1$) using single-source  likelihood given
%by Eq.~(2). We  found that the calculated  distribution  of test statistic $\mathrm{TS}_{j}$ was compatible with the  null hypothesis. This result  confirmed our  assumption that such combination of 
%flares cannot be detected  at a $5\sigma$ level  by at least a time-integrated method.}.    

Figure~\ref{figure7} (A) and (B) show that  also in this case  the method 
can recover the true  values of the   spectral index ($\gamma=2$),  the    mean   number of injected signal events (8)  and the total flare duration (28 days). 
In addition,    the  overall  signal injection can be decomposed into  about 8  data segments (doublets). %Note that by construction, they partially overlap.

Sorting these data segments  in time  we  obtain   the   distribution {\it   of individual flares}
in time  as  shown in Figure~\ref{figure7} (C). Three flares can be clearly distinguished,  
with a duration of  about 5 days, 5 days and 10 days, respectively.

The average value of the  test statistic  and the average number of  signal events   from
the best fit  (of  the likelihood  $\tilde \mathcal{L}$)  is  presented in Figure~\ref{figure7}~(D)~and~(E). 
We observe a similar behavior as  in  the Example~2.  
However,  due to  a smaller   time  gap  between   individual  flares, a factor 4 smaller
than for the  flares studied in  the Example~2, the structures  are less   pronounced.
 Note  that the distributions presented in Figure~\ref{figure7}~(D) and (E) start  to saturate   
at  a point   corresponding to the overall period of enhancement emission  (above 28 days). 

Finally, in Figure~\ref{figure7} (F)  the  average spectral
 index $\left< \hat\gamma_s(m) \right>$ as a function of time is shown. The  source 
spectral index has a value of about 2.2 and  differs by about  10\% from the true value ($\gamma_{s}=2$). 
Fluctuations  of the spectral index  also increase  for  times  greater 
than  the total flare duration (above 28 days).

This example demonstrates that the proposed   algorithm  can  find  a few  weak flares separated in time,
 which cannot be found by other standard methods~\cite{Braun2}.   
  
\section{Performance of the method}
% In this section we  compare  the discovery potential of the method which we  propose with what 
%  presented in ~[\cite{Braun2}]. The  result comparison
% is  rather  qualitative, because   a  precise  comparison  would require   the same  simulation conditions i.e. for example  the same  data set,  the same   background  level etc.  Here 

In this section we calculate  the number of events needed for discovery (i.e. the discovery potential)
for the cases of  Example~1, 2 and~3 and for  different flares durations $\Delta T$ and  overall data periods $\Delta T_{\mathrm{data}}$.
 The discovery potential   is defined as the average number of signal events required to achieve a p-value less than $2.87\times10^{-7}$ (one-sided $5\sigma$) in 50\%  of the trials. The  comparison  below  gives us  an idea about the  performance of the  proposed method and its limitations. 

\begin{figure*}[ht]
 \centering
 \includegraphics[width=0.48\textwidth,height=5cm]{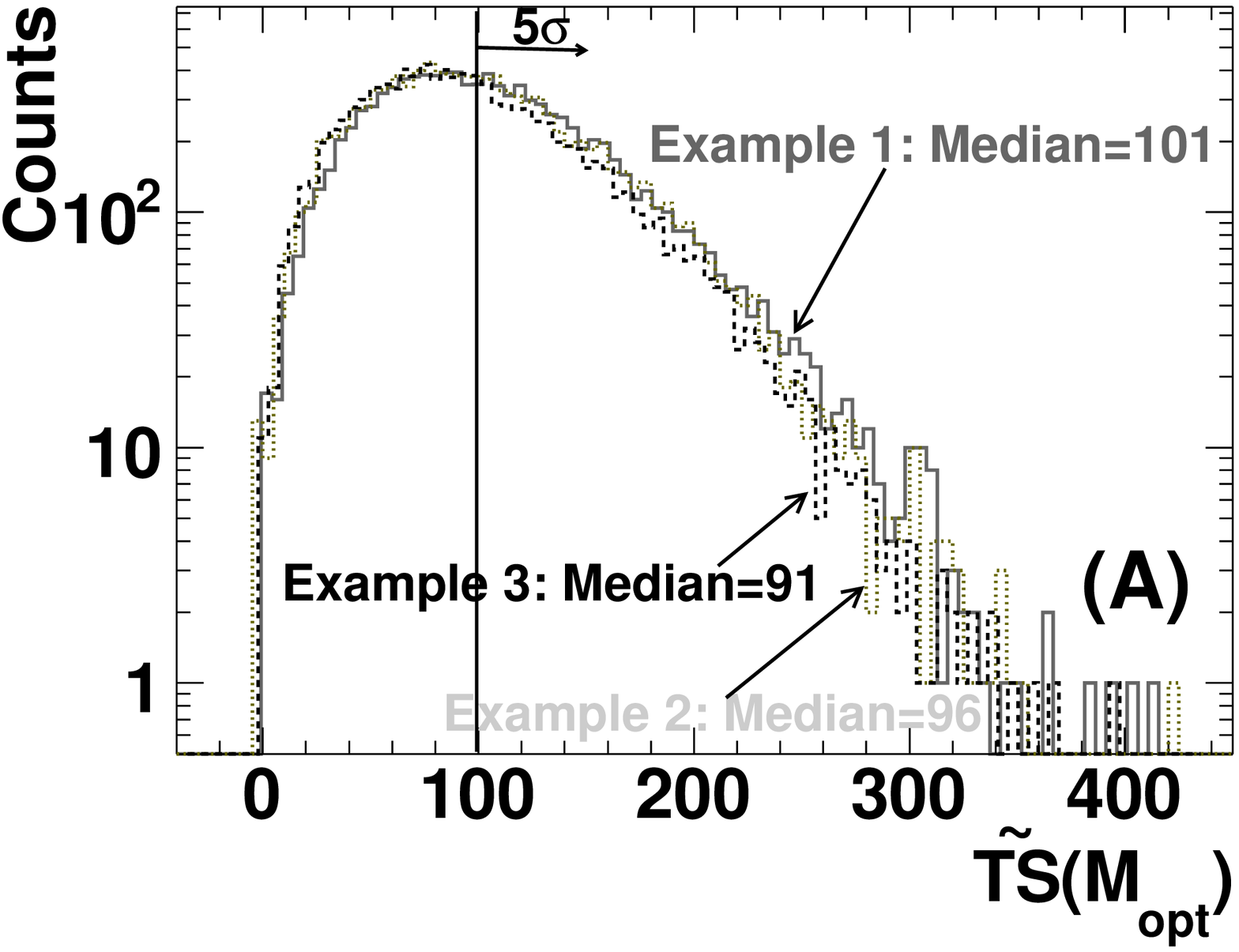}
 \includegraphics[width=0.48\textwidth,height=5cm]{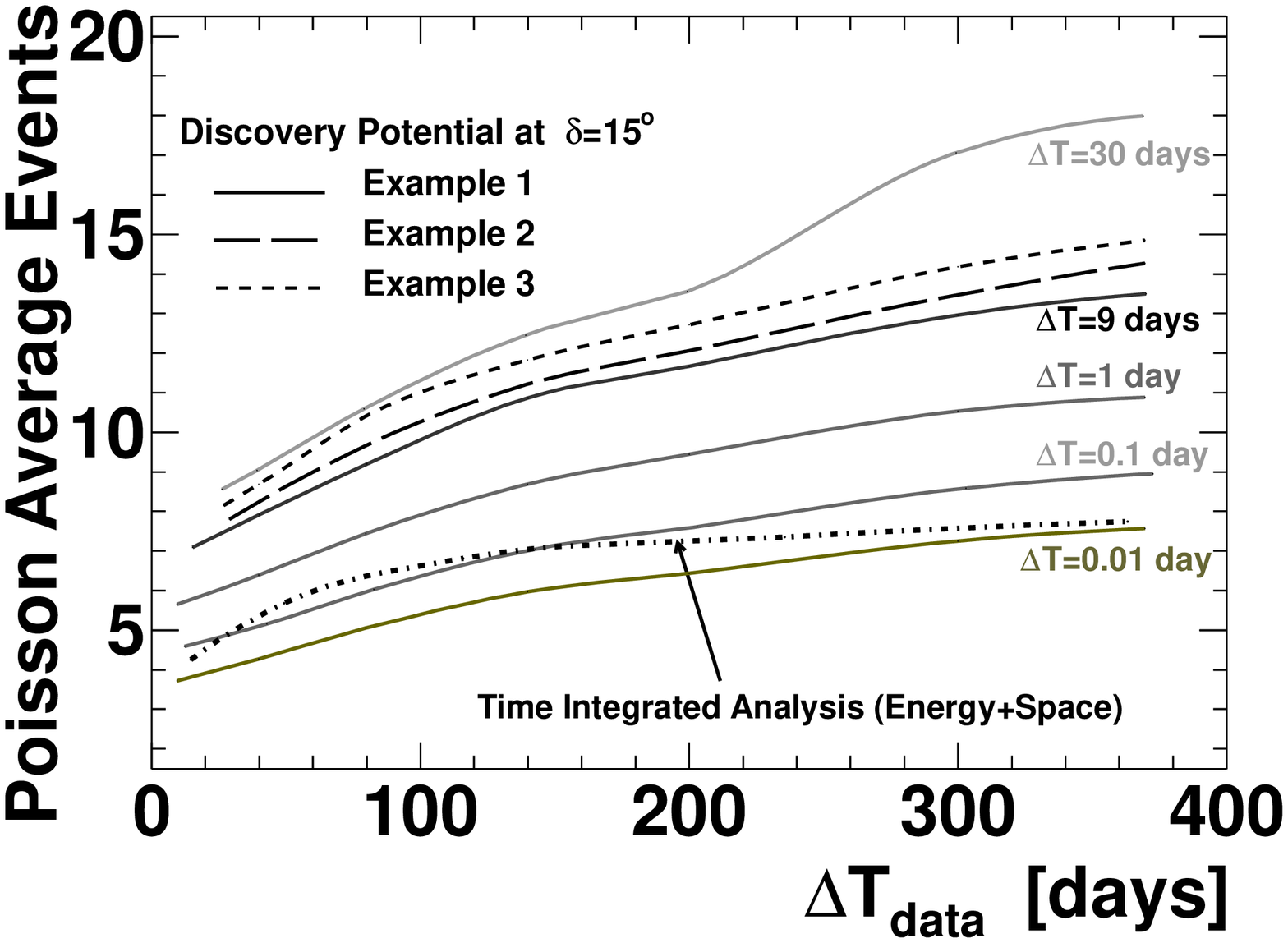}
 \caption{\small (A) Distribution of the  test statistic $\widetilde{\mathrm{TS}}(M_{\mathrm{opt}})$ for  background
plus  signal  simulations with a Poisson mean   of  8   signal events. The  results  obtained
for   one flare  with duration  9 days (Example~1), two flares with a total  duration (including gaps)
of 35 days (Example~2)  and three weak flares with a  total duration of  28 days (Example~3) are compared. 
The vertical  dashed line  indicates the 5$\sigma$ threshold for background only case;
(B) Discovery potential   as a function of an overall data period  $\Delta T_{\mathrm{data}}$ for  one flare  with different durations and for two flares and three weak flares.      
The dashed line shows results of a  time-integrated analysis  performed in this work with  an energy term in the PDF.}\label{figure9}
 \end{figure*}

In Figure~\ref{figure8} we  show  the distribution 
of the global test statistic   $\widetilde{\mathrm{TS}}(M_{\mathrm{opt}})$ for MC background-only simulations (A)   
and  for MC background-plus-signal simulations (B),  with  signal events distributed  according to a Poisson  distribution
 with mean 4, 8 and  12 (the case of   Example~1: one flare with duration 9 days).
From   Figure~\ref{figure8}~(A) we can estimate the  $\widetilde{\mathrm{TS}}|_{5\sigma}(M_{\mathrm{opt}})$ threshold for a $5\sigma $ execss in the background  only  case, 
which is 101. The $5\sigma$  threshold was calculated for a band of  $\pm6^{\circ}$~\footnote{Calculations with a larger band ( $\pm10^{\circ}$)
 does not  show changes in  the $5\sigma$ threshold.} centered at declination $15^{\circ}$ and for an overall  data period of $\Delta T_{\mathrm{data}}=40$ days.  From  Figure~\ref{figure8}~(B) we can see  that  this threshold  is passed in 50\% of trials
 when the average number of added signal events is 8.  The  discovery potential is therefore  about 8 events. 

In Figure~\ref{figure9}~(A)   the global test statistic  distribution
 for the  case of    Example~1, 2~and ~3 is shown. 
The median  is 101, 96 and 91, respectively, which  means that the number of events required  for discovery
(assuming a threshold  of 101) is $8.0$, $8.4$ and $8.9$. 
 Note that we have a similar  number of events required for discovery for Example 2 and Example 3
 in which signal events are injected over a few flares, but  such cases cannot be tested  with   the method  presented in~\cite{Braun2}.

In Figure~\ref{figure9}~(B) the discovery potential as a function of different overall data periods is shown.
The    number of events  required for discovery  as  expected  depends on the overall  data period $\Delta T_{\mathrm{data}}$ considered. 
This is because the number  of  signal-like events increases with time 
 and therefore  also    leads  to  a  higher $5 \sigma$  threshold.  For example,
for an overall   data period  of $40$ days   the average  number of  signal-like events  is 5, while
it is 43  for a data period of  $365$ days. Such changes in the number of signal-like events 
 lead to an increase of the  $5\sigma$  threshold  from  101 to 180. As a consequence
 the number of events needed for discovery increases by   about 63\%,  from 8.0  up to 13  
for a  flare with a duration  of 9 days.
Comparing with the   results  presented  in~\cite{Braun2} where we need about 9 events for discovery 
  for a method using unknown burst time with energy PDF and one year data period (Figure~4 in~\cite{Braun2}),   we  see that  our method  requires   about  44\%  more  events for discovery in case all events are injected in one single flare (Example 1). This is because our algorithm stacks also background fluctuations,  and thus leads  to   a  higher $5 \sigma$  threshold than the  threshold obtained by a single-source likelihood based method. However, it must be noticed that this comparison is only qualitative, since the signal-to-background ratio, which reflects the detector efficiency and signal properties used in both simulations, is different in the two cases. A more quantitative comparison of both  two methods
 has to be achieved on the same simulated data set.
As we also  expect, the  number of events needed for discovery  
 decreases,  when we consider  flares with shorter duration.
 For example,  for a flare with duration of 30 days, 1 day,  0.1 day and 0.01 day   
in average  about  9,  6, 5 and 4   events are needed for  discovery  within  
 $\Delta T_{\mathrm{data}}=40$ days. Note that in this case 
 for  flares with shorter  duration (below 0.1 days) the number of events is smaller than 
the number of events for a time-integrated analysis,  see Figure~\ref{figure9}~(B) for the 
 method labeled  ``Time Integrated Analysis (Energy+Space)''. 

In Figure~\ref{figure9}~(B) the results of calculations for conditions of  Example 2   and 3  are  also  shown. In general,  we see a similar trend  i.e. the number of  events needed for discovery  increases  when a larger overall  data  period is studied.  It  is also  seen
 that   for Example~2 the number of events for discovery is smaller than for a single flare   
with similar duration.  However,   care  must be taken in
  such a comparison because,  as we already  pointed out before, 
 multiple   combinations of individual flares cannot be  found  by the  standard
 point source search  algorithm as proposed in~\cite{Braun2}. This is especially true for Example~3.
  In  other words,    the performance of the method  improves when we study more flares  
distributed in time,  especially if they are weak. 

Up to now we  also do not   exploit  the fact   that  the  algorithm  provides us information 
about  the number of individual flares distributed in time  or even about the number of 
data segments which compose one flare or  a few individual flares separated in time. 
Using  such  information,  we can calculate  a   better sensitivity.  
By dividing  the sensitivity  by the number of flares or 
 data segments  $M_{\mathrm{opt}}$, we can calculate  a sensitivity  per  individual flare
or even  per  segment. For   examples presented here   (which correspond to signal-like doublets)
  the number of $M_{\mathrm{opt}}$ is about 8  and the  number of individual flares is larger than 1, 
hence the sensitivity    can  be improved by   about the same factor.

\section{Summary \& Conclusion}
We have presented a  method to  search for neutrino 
flares from point sources without an a priori assumed time structure.
The method  considers  only data segments which contain   signal-like  doublets, 
and uses  a  test-statistic  term as  their weights in a stacking-like  calculation 
for the global maximum likelihood. We  have shown that this  method 
 can recover the  true values of the source spectral index, the   flare duration,
and the total  number  of  injected signal events within uncertainties not larger than 10\%.
In addition,  our  algorithm  provides relevant physics  information about  the distribution of flares 
in time  and  their internal structures. This information   can be used to calculate 
a sensitivity  per  individual flare, which is  usually better than the sensitivity obtained from the other methods.

For  standard  cases (one ``strong``  flare), the discovery potential of  the method 
  is  about  44\%    worse than  the standard point source analysis with unknown
 duration of the flare.  This is  because our  method  stacks  all significant  background  fluctuations in a given period of data and leads to a higher  threshold for discovery.  
However even  in  such  a case,  the number of events required   for discovery is   smaller compared to  a time-integrated search as  soon as
 the flare  duration is  less than a few hours. 
 When the number of  individual flares   analyzed  is  increased  the number of events needed for discovery decreases,
 especially in the case of    a few weak  flares  distributed  over a longer  period (from a few days up to 100 days). 
Such cases  of a few weak flares cannot be discovered  by  the standard  point search algorithm~\cite{Braun2}.  
% In addition,  our  algorithm  provides relevant physics  information about  the distribution of flares 
% in time  and  their internal structures. This information   can be used to calculate 
% sensitivity  per  individual flare, which are usually better than the sensitivity obtained from the other methods. Moreover they provide
% fundamental input for a  phenomenological interpretation, when compared to multiwavelength data. 

\section{Acknowledgments} We would like to thank A. Kappes and C. Spiering for useful discussions.
We acknowledge the support from  the Young Investigators Program of the Helmholtz Association. 
% {\it we should also  thanks  Alexander  and Christian}

\bibliographystyle{elsarticle-num}

\end{document}